\documentclass[final,3p,times,a4]{elsarticle}
\pdfoutput=1 % if your are submitting a pdflatex (i.e. if you have
\usepackage{color}
\usepackage{graphicx}
\usepackage[utf8]{inputenc}
\usepackage{natbib}
\usepackage{amsmath, amssymb}
\usepackage{mathtools}
\usepackage{url}
\usepackage{wasysym}
\usepackage{float}
\usepackage{gensymb}
\usepackage{textcomp}
\usepackage{stmaryrd}
\usepackage{todonotes}
\usepackage{subfigure}
\usepackage{color}

\usepackage{hyperref}

\definecolor{darkgreen}{rgb}{0,0.7,0}

\newcommand{\overbar}[1]{\mkern 1.5mu\overline{\mkern-1.5mu#1\mkern-1.5mu}\mkern 1.5mu}
\newcommand{\Lagr}{\mathcal{L}}

\begin{document}

\hfill{\small TTK-18-15}

\begin{frontmatter}

\title{MontePython 3: boosted MCMC sampler and other features}
\author{Thejs Brinckmann \& Julien Lesgourgues}
\address{Institute for Theoretical Particle Physics and Cosmology (TTK), RWTH Aachen University, Otto-Blumenthal-Strasse, 52057, Aachen, Germany.}
\address{E-mail: \href{mailto:brinckmann@physik.rwth-aachen.de}{brinckmann@physik.rwth-aachen.de},\href{mailto:lesgourg@physik.rwth-aachen.de}{lesgourg@physik.rwth-aachen.de}}
\journal{Physics of the Dark Universe}

\begin{abstract}
MontePython is a parameter inference package for cosmology. We present the latest development of the code over the past couple of years. We explain, in particular, two new ingredients both contributing to improve the performance of Metropolis-Hastings sampling: an adaptation algorithm for the jumping factor, and a calculation of the inverse Fisher matrix, which can be used as a proposal density. We present several examples to show that these features speed up convergence and can save many hundreds of CPU-hours in the case of difficult runs, with a poor prior knowledge of the covariance matrix. We also summarise all the functionalities of MontePython in the current release, including new likelihoods and plotting options.
\end{abstract}

\end{frontmatter}

%\tableofcontents
\newpage
\section{Introduction}
\texttt{MontePython} \citep{Audren:2012wb} is an MCMC sampling package in \texttt{Python} used for parameter inference in cosmology, similar to \texttt{CosmoMC} \citep{Lewis:2002ah,Lewis:2013hha} and \texttt{CosmoSIS} \citep{Zuntz:2014csq}. The modular nature of \texttt{MontePython} means modification of the code is particularly easy, and encourages implementation of specific modules to other Python sampling packages, e.g. the extensive library of cosmological likelihoods\footnote{A practice the authors fully support and encourage, with proper citations and credits.}.  \texttt{MontePython} has two different modes: when running with

\texttt{> python montepython/MontePython run <options>}

\noindent it is a sampler (similar to \texttt{CosmoMC}), and when running with

\texttt{> python montepython/MontePython info <options>}

\noindent it is a tool for analyzing MCMC chains and plotting results (similar to \texttt{GetDist}\footnote{\url{http://getdist.readthedocs.io/en/latest/}}).

The code is currently interfaced with the Boltzmann code \texttt{CLASS}~\cite{Lesgourgues:2011re,Blas:2011rf,Lesgourgues:2011rg,Lesgourgues:2011rh} and extensions thereof, e.g. \texttt{HiCLASS} \citep{Zumalacarregui:2016pph} and \texttt{SONG} \citep{pettinari:2013a}. There also exist some publicly available branches of CLASS achieving different purposes, e.g.
\texttt{ExoCLASS} for advanced energy injection, recombination and reionisation features \citep{Stocker:2018avm},
\texttt{CLASS\_SZ} for Sunyaev-Zel'dovich observables \citep{Bolliet:2017lha},
\texttt{CLASSgal} for computing the number count $C_\ell$'s \citep{DiDio:2013bqa}\footnote{this feature has also be implemented later in the main \texttt{CLASS}, but with small differences in the two implementations. The original \texttt{CLASSgal} code is still available at \url{https://cosmology.unige.ch/content/classgal}}, or a branch incorporating nonlocal contributions to General Relativity \citep{Dirian:2016puz}\footnote{see the pull request \#86 in \url{https://github.com/lesgourg/class_public}}.

 In principle, it could easily be extended for use with e.g. \texttt{CAMB} \citep{Lewis:1999bs,Howlett:2012mh}, via the new Python wrapper\footnote{\url{http://camb.readthedocs.io/en/latest/}}, or PyCosmo \citep{Refregier:2017seh}, a Boltzmann code in Python.
  
In this paper, we present the latest development of \texttt{MontePython} over the past couple of years. In particular, we introduce two new ingredients that both contribute towards improving the performance of Metropolis-Hastings sampling.
In Section \ref{sec:sampling}, after recalling the way in which the Metropolis-Hastings algorithm is implemented in \texttt{MontePython}, we present 
a new adaptation algorithm for the jumping factor. 
In Section \ref{sec:fisher}, we detail our strategy for calculating the  Fisher matrix and its inverse, which can be used as a proposal density for a Metropolis-Hastings run. 
In Section \ref{sec:performance}, we provide several examples of runs showing that these features speed up convergence and can save many hundreds of CPU-hours in the case of difficult runs, with a poor prior knowledge of the covariance matrix. 
In the various appendices, we summarise all the functionalities of \texttt{MontePython} in the current release, including extended cosmological parameter definitions with respect to \texttt{CLASS} in \ref{sec:parametrizations}, sampling options in \ref{sec:sampling_options}, analysis and plotting options in \ref{sec:plotting_options} and likelihoods in \ref{sec:likelihoods}. Indeed, the new release of the code incorporates several new plotting options and even more new likelihoods based either on current or mock data.

The version of the code described in this paper has version number 3.0 and is available at \url{https://github.com/brinckmann/montepython\_public}
\section{Metropolis-Hastings sampling strategy}
\label{sec:sampling}
\texttt{MontePython} can switch between different ways to explore parameter space, which include Metropolis-Hastings, Nested Sampling, Cosmo Hammer, and a new Fisher sampling method described in section~\ref{sec:fisher}.  These different algorithms are called \texttt{methods} in the code, and the same list of methods also includes post-processing algorithms like Importance Sampling or Adding Derived Parameter(s)\footnote{The full list of methods can be viewed with {\tt MontePython.py --run --help}, and is of the form \texttt{-m \{MH, NS, CH, IS, Der, Fisher\}.}}.

The default method is the Metropolis-Hastings algorithm, working since {\tt v2.0.0} (2013) with a {\it fast sampling} method quickly summarised in section~\ref{sec:fast_sampling},  and since {\tt v2.2.0} (October 2015) with a covariance matrix update method 
summarised in section~\ref{sec:superupdate}. In this release {\tt v3.0.0} we extend the latter to update also the jumping factor, as described in section~\ref{sec:superupdate}, and we call the new approach {\tt superupdate}.

\subsection{Fast sampling}
\label{sec:fast_sampling}

In \texttt{MontePython} the Metropolis-Hastings draws random jumps in parameter space from a Gaussian proposal density. The latter is encoded in a matrix $\mathbf{C}$, describing the parameter correlations and the standard deviations relative to each other, and an overall jumping parameter $c$, such that the parameter jumps $\Delta \mathbf p$ are generated randomly from the probability distribution ${\cal P}={\cal N} \exp( -\frac{1}{2 c}\Delta \mathbf{p}^T \mathbf{C}^{-1}\Delta \mathbf{p})$. Thus the actual covariance matrix of the proposal density is $c \mathbf{C}$. The standard way to generate random vectors from a multivariate Gaussian probability is to go to a basis of independent parameters, like the basis of the eigenvectors of $\mathbf{C}$; to generate independent random displacements along the eigenvectors; and to project back to the original space. 

It is well-known that optimal proposal densities generate an acceptance rate of the order of 0.25, and that for Gaussian posterior distributions, this can be achieved when $\mathbf{C}$ is a good approximation to the covariance matrix of the {\it posterior distribution}, while the jumping parameter is fixed to $(2.4)^2$ \citep{Dunkley:2004sv}.  Note that this jumping parameter applies when generating one single random number and moving in one single direction. Alternatively, for each jump, one can generate $N$ random numbers and move in $N$ directions simultaneously, but then each of these $N$ random numbers should be drawn from a Gaussian distribution with variance $c=j^2/N$, where $j=2.4$ is called the {\it jumping factor}.

Thus, in absence of fast sampling,  the jumps could just be generated each time independently and randomly from ${\cal P}$ (global method), or in cycles of $N$ draws along each of the $N$ eigenvectors of $\mathbf{C}$ (sequential method). These methods can still be activated\footnote{with the {\it jumping} flag {\tt -j \{global, sequential\}} instead of the default {\tt -j fast}.} in \texttt{MontePython}, but they are sub-optimal in presence of likelihoods with nuisance parameters.

{\it Fast sampling} was proposed by \cite{Lewis:2013hha} for MCMC parameter estimation. For this sampling method, we separate the sampling of fast nuisance and slow cosmological parameters to optimize performance when dealing with a large number $N_\mathrm{fast}$ of nuisance parameters. 

Generating displacements along the eigenvectors mixes slow and fast parameters and does not allow for high-speed explorations of the fast parameter space only. But introducing eigenvectors is not the only way to go to a parameter basis in which the proposal density is orthogonal. In particular, one can perform a Cholesky decomposition of the covariance matrix into $\mathbf{C} = \mathbf{L} \mathbf{L}^T$ where $\mathbf{L}$ is a lower triangular matrix. In the space  of the vectors $\Delta \mathbf{p}'$ related to the physical parameters through
$\Delta \mathbf{p}' = \mathbf{L}^{-1} \Delta \mathbf{p}$, the proposal density is orthogonal, so the jumps can easily be generated by drawing random numbers for each component of $\Delta \mathbf{p}'$ with a single one-dimensional Gaussian probability distribution of variance $c$, and projecting back to $\Delta \mathbf{p} = \mathbf{L} \,  \Delta \mathbf{p}'$. The great advantage over the previous eigenvector-based scheme is that 
when $\Delta \mathbf{p}'$ only has non-zero components above a given index, this is true also for $\Delta \mathbf{p}$.  Thus one can generate some jumps that will leave the slow parameters unchanged.

We begin by ordering our input parameters in blocks according to computational time. In practise, this is simply achieved by writing them in the right order in the input parameter file.  The first block is that 
of cosmological parameters requiring new calls to the Boltzmann code. The next blocks are nuisance parameters for a given likelihood, which can be changed without requiring a Boltzmann code evaluation if the cosmological parameters are held fixed. The nuisance parameter blocks should be ordered from the slowest to fastest likelihood. When a nuisance parameters is common to several likelihoods, it should just be declared within the slowest block. We call $M$ the number of blocks and $d_j$ the number of parameters in the $j$-th block, with $d_1 = N_\mathrm{slow}$ being the number of cosmological parameters. \texttt{MontePython} will automatically detect the number $M$ of blocks and will expect the user to pass an $M$-dimensional 
over-sampling vector $\mathbf{F}$. $F_1$ is the over-sampling factor of the cosmological parameter and is normally fixed to one. The other entries are the required number of redundent sampling for each of the other blocks. 

When running chains, for $j=1,...,M$, we generate sequences of $F_j d_j$ random jumps in the $d_j$ components of $\Delta \mathbf{p}'$ corresponding to the $j$-th block. In other words, during $F_j d_j$ steps, we generate $d_j$ random numbers for each of the relevant components of $\Delta \mathbf{p}'$, drawn from a Gaussian distribution with standard variance $c=j^2/d_j$.
Thus each full cycle consists  of $\sum_j F_j d_j$ random jumps, with only $F_1 d_1= N_s$ of them requiring a call to the Boltzmann code. Later we will call this number the Fast Parameter Multiplier (FPM):
\begin{equation}
\mathrm{FPM} = \sum_j F_j d_j.
\label{eq:FPM}
\end{equation}
There is no precise rule to fix the over-sampling factors $F_{2,...,N}$. These factors should be increased for faster likelihoods and/or larger numbers of nuisance parameters in the block. With too low numbers, one would not enjoy the advantages of the slow-fast parameter decomposition. With too high numbers, the time spent in the $(\sum_j F_j d_j-N_s)$ iterations over fast parameters could be significant compared to the time spent in the $N_s$ iterations over slow parameters, and the convergence of the results for the cosmological parameter would be delayed. For instance, for the nuisance parameters of the Planck likelihood, we usually apply an oversampling factor of 4.

\subsection{Update and Superupdate}
\label{sec:update_and_superupdate}

While the Metropolis-Hastings algorithm would in principle require Markov Chains, i.e. chains with a proposal density that is constant in time, it is highly desirable to implement some automatic update algorithms in order to get converged results even when starting from bad guesses for the Gaussian proposal density. We recall that the propoosal density is parametrised as ${\cal P}={\cal N} \exp( - \frac{1}{2c} \Delta \mathbf{p}^T \mathbf{C}^{-1} \Delta \mathbf{p})$ and thus depends on two quantitites, the covariance matrix $\mathbf{C}$ and
the jumping parameter $c$ related to the jumping factor $j$. \texttt{MontePython} has two complementary options for speeding up the convergence of Metropolis-Hastings runs:
\begin{itemize}
	\item \texttt{--update} $U$: update of covariance matrix, $\mathbf{C}$, every $U$ cycles [default: $U$ = 50]
	\item \texttt{--superupdate} $S\!U$: additionally, update of jumping factor, $j$, starting $S\!U$ cycles after each covariance matrix update [default: $S\!U = 0$, meaning ``deactivated''; recommended: 20]
\end{itemize} 
Once certain criteria are met, the covariance matrix will be updated periodically and the jumping factor will be adapted every step. This leads to dramatic improvements in runtime, especially for runs with little prior knowledge in the form of an appropriate starting stepsize or a good initial proposal distribution. In the next two sections we describe the strategy chosen for these two schemes.

\vspace{0.5cm}

\subsection{Update strategy}
\label{sec:update}

The covariance matrix update mechanism was implemented early on in \texttt{CosmoMC} \cite{}, and is part of \texttt{MontePython} since {\tt v2.2.0} (October 2015), through the flag \texttt{--update}. The \texttt{MontePython} and \texttt{CosmoMC} implementations of this feature are very similar. For instance, in both codes, the decision to start or stop the update mechanism depends on the value of the Gelman-Rubin statistic, $R$ \citep{Gelman:1992zz}, for the most poorly converged parameter.
The update starts when the number $\mathrm{max}(R-1)$ computed from the second half of each chain goes below 3, and stops when it goes below 0.4. 
The difference between the two implementations only resides in two aspects:
\begin{itemize}
\item {\it Non-MPI-user friendliness.} The most straightforward way to launch multiple chains is to run \texttt{MontePython} with \texttt{MPI}, e.g. for 8 chains:
\texttt{mpirun -np 8 python montepython/MontePython.py run ...}  \\
The alternative would be to launch 8 chains manually, or within a small shell script with a \texttt{for} loop. With \texttt{CosmoMC}, this second option would be incompatible with the covariance matrix updating. In \texttt{MontePython}, because installing \texttt{MPI} can sometimes be cumbersome, we chose to code the \texttt{--update} mechanism in such a way that it will work equally well with or without \texttt{MPI}. In the latter case, the user should just run several times \texttt{python montepython/MontePython.py run ...}, and the update mechanism will still start and use the information from {\it all} the chains running in the same directory\footnote{There is no need to know how this is implemented in the code, because it is fully transparent to users. In brief, the key point is that \texttt{MontePython} reads and writes the covariance matrix in a file, rather than using the Message-Passing-Interface. The only difference between MPI and non-MPI runs is that in the former case, the code can define a ``master chain'' and some ``slave chains'', and only the master chain occasionally pauses in order to update the covariance matrix; in the latter case, all chains occasionally pause for the same purpose, but this does not affect convergence, and it only increases the total running time by a very small amount. Note that exchanging information on the covariance matrix through a file could have a potential inconvenience: When the user analyses an on-going run with the  \texttt{info} mode, she/he could generate a new covariance matrix that may interfere with the automatic updating mechanism. This is not the case because when the user runs in \texttt{info} mode, the covariance matrix calculation is de-activated by default; it is only activated with the  \texttt{--want-covmat} flag.}.
\item {\it Keeping only Markovian steps in the final results.} To be rigorous, the user would like to base his final results and plots on true Markovian chains. This is what happens by default with \texttt{MontePython}. Indeed, every time that the covariance matrix is updated, the code writes in all chains files a comment line starting with\\ \texttt{\# After <k> accepted steps: update proposal...}\\ which also contains information on the current convergence (thus these comments can also be used to scrutinise what is happening with the run). When analysing the chains with the \texttt{info} mode, by default, the code will only consider the part of the chains after the last update, i.e. the Markovian part. If the user wants to de-activate this behaviour in order to get more points in the chains, she/he can use the flag \texttt{--keep-non-markovian}.
\end{itemize}
Finally, the update periodicity is controlled by the $U$ input parameter (default: $U=50$), which is in units of cycles. Given that a cycle 
consists in FPM steps (see eq.~(\ref{eq:FPM})), the update takes place every 
\begin{equation}
N_{\rm update} = \rm U \times \rm FPM
\end{equation}
steps (here we are referring to proposed steps, not accepted steps).

\subsection{Superupdate strategy}
\label{sec:superupdate}

The covariance matrix updating does not fully achieve the task of reaching optimal convergence conditions automatically. The other part of the proposal density is the jumping parameter $c$, related to the jumping factor $j$. If $j$ is too large, the acceptance rate ($a.r.$) is too small and the number of accepted models remains insufficient to extract statistical information. If $j$ is too small, the $a.r.$ may get close to one. In that situation the chains would grow rapidly, but adjacent points would be very correlated, and the chains would not necessarily sample the full posterior distribution. Thus one should target a compromise value of the acceptance rate. Since the work of \cite{Dunkley:2004sv}, cosmologists usually aim at $a.r.\simeq 0.25$ (although larger values in the range 0.3-0.5 would in principle still be acceptable).

While $f=2.4$ matches this goal for a multivariate Gaussian posterior, many runs target non-Gaussian posteriors, e.g. due 
to non-trivial priors on cosmological parameters (like the requirement of a positive neutrino mass) or to strongly non-Gaussian posteriors for the nuisance parameters (like those of the Planck high-$\ell$ likelihoods). The current practise consists in training a bit and trying manually different values of $f$ until the acceptance rate is correct. For instance, one quickly comes to know that, e.g., a given version of the Planck likelihood usually needs a given value of $f$ to achieve $a.r.\simeq 0.25$. Of course, it would be better to let the code find this value automatically for each combination of a model and a dataset. An automatic jumping factor adaptation would also make the code more powerful when starting from a very bad proposal density (e.g., when adding many new free parameters to a previous run, or when investigating a very constraining set of likelihoods when only the covariance matrix of a much less constraining set is available).

For that purpose, we added to version $\texttt{v3.0.0}$ the new option \texttt{--superupdate}, which is complementary to \texttt{--update}: they should normally be used in combination, but then it is only necessary to pass the \texttt{--superupdate} flag since \texttt{--update} is activated by default. 

Note that other schemes to update the full proposal density (rather than just the covariance matrix) have been investigated in the past. For instance, an Adaptative Metropolis algorithm for single-chain runs was proposed in \cite{haario2001}, and a version of this algorithm was implemented in a cosmology MCMC code by the CAMEL collaboration \citep{Henrot-Versille:2016htt}. After trying this method for single chains, we adapted it freely to the multi-chain case, in a way which remains compatible with the traditional covariance matrix update scheme (\texttt{--update}). \\

\noindent {\it Overall strategy.}
When running with \texttt{--superupdate} $S\!U$, the code starts from a jumping factor that can be set manually with a command flag (e.g. \texttt{--f 2.2}), but when nothing is passed,  2.4 is used by default. The code keeps a record of the acceptance rate, $a.r.$, and of the jumping parameter, $c$, of the last $S\!U$ cycles, i.e. of the last $S\!U \times \rm FPM$ steps. It also keeps track of the average $\overbar{a.r.}$ and $\bar{c}$ over these last $S\!U$ cycles.  This information is used to compute the start- and stop-criteria. Since \texttt{--superupdate} requires \texttt{--update} to be active, the run can be divided in several ``update sequences'', which are the ensemble of steps between two consecutive updates. The first ``update sequence'' is just the time until the first update. The basic principle of superupdate is to adapt the jumping factor at each step according to the recurrence relation
\begin{align}
c_{k} = c_{k-1} + \frac{1}{(k-k_{\rm update})}(\overbar{a.r.} - 0.26) \ ,
\label{eq:recursive1}
\end{align}
where $k$ is the current step number, while $k_{\rm update}$ is the first step number of each new ``update sequence''. For the first sequence, $k_{\rm update}=0$. This recurrence relation leads to faster updating at the beginning of each new sequence, and to slower updating and safe convergence properties after some time.\\

\noindent {\it Starting the jumping factor update.}
The code starts applying the recursion relation~(\ref{eq:recursive1}) when two conditions are met:
\begin{itemize}
\item We do not want to update the proposal distribution too early, as it could be based on chains still in the burn-in phase. For this reason, we wait until the chains have reached a certain level of convergence: the numbers $(R-1)$ computed from the second half of each chain should be below 10 for all parameters.
\item We wait until we have done $S\!U$ cycles since the beginning of the new ``update sequence'', or since the very beginning if we are still in the first sequence: $(k-k_{\rm update})\geq S\!U \times FPM$. Since the mean acceptance rate $\overbar{a.r.}$ is computed over the last $S\!U$ cycles, the recurrence will only take into account some steps from the same ``update sequence''. Choosing $S\!U \gtrsim 20$ ensures that the mean acceptance rate is not computed over too small a sample, where shot noise may lead to an acceptance rate significantly different from the target one (i.e. due to random fluctuations in the acceptance rate leading to prematurely stopping adaptation of the jumping factor).  We recommend using $S\!U = 20$, as we found this to be a good compromise between efficiency and precision, but higher values can be considered in order to decrease the impact of superupdate (the jumping factor would start evolving later and would perform smaller excursions) or to further decrease the impact of shot noise on the determination of the jumping factor. Note that with $S\!U>U$, superupdate would sometimes only be active after the very last update of the covariance matrix, in the final stage of the run, when the convergence is already good (or possibly before the chains are well enough converged for updating the covariance matrix to begin). Thus one should normally consider the range $20 \leq S\!U < U$ only.
\end{itemize}

\noindent {\it Rescaling when the covariance matrix gets updated.}
Since the true covariance matrix of the Gaussian proposal density is in fact given by the product $c\mathbf{C}$, it would be sub-optimal to leave $c$ unchanged when the matrix $\mathbf{C}$ is updated at the beginning of each new ``update cycle''. Suppose for instance that in the $n$-th ``update cycle'', a good jumping factor $c_n$ has been found in combination with a covariance matrix $\mathbf{C}_n$ (by this we mean that the acceptance rate has the correct order of magnitude). If at the beginning of the next cycle the matrix is adapted to a $\mathbf{C}_{n+1}$ which is much smaller, while $c_{n+1}$ restarts from the same value $c_n$, then obviously the whole proposal density will shrink and the acceptance rate will increase too much. We can limit this effect by requiring analytically that at each covariance matrix update, the volume probed by the full proposal density remains constant, which is achieved simply by imposing:
\begin{align}
c_{\rm after}^N \det(C_{\rm after}) = c_{\rm before}^N \det(C_{\rm before}) \ ,
\end{align}
where $N$ is the number of free (slow+fast) parameters. This rescaling might not be very efficient when the evolution of the covariance matrix comes from only one or few parameters, but in general, it is the best simple guess that one can do. In terms of the jumping factor, this gives:
\begin{align}
j_{\rm after} = j_{\rm before} \left[\dfrac{\det(C_{\rm before})}{\det(C_{\rm after})}\right]^{\dfrac{1}{2 N}} \ .
\end{align}
Note that, for the first update of the covariance matrix, the logic behind this re-scaling does not hold. Indeed, if we started from a poor input covariance matrix, the first re-scaling of the jumping factor may be completely unrealistic. For safety, at the first update time, we reset the jumping factor to the input value (provided via \texttt{--f} [default: 2.4]). \\

\noindent {\it Stopping the jumping factor update.} 
We adapt the jumping parameter until three conditions are met:
\begin{itemize}
\item The $a.r.$ should converge to 26\% with a tolerance of 1 percent point\footnote{This behaviour is controllable by the options \texttt{--superupdate-ar} and \texttt{--superupdate-ar-tol}.} (in many cases, the $a.r.$ starts low and increases to the optimal value, then  the adaptation will stop when the code reaches 25\%):
\begin{align}
|\overbar{a.r.} - 0.26| < 0.01 \ .
\end{align}
\item In addition to the $a.r.$ criterium, in order to stop adaptation of the jumping parameter we also require that it is stable,
\begin{align}
\left| \dfrac{\bar{c}}{c_{k-1}} -1 \right| < 0.01 \ ,
\end{align}
where $\bar{c}$ is the mean of the jumping parameter over the last $S\!U \times \rm FPM$ steps. Also, we do not wish to allow the jumping parameter to converge to arbitrarily low values, as the risk of chains getting stuck in local minima would increase. Therefore, we introduce a minimum for the jumping factor corresponding to 10\% of the initial jumping factor. If a small jumping parameter is desired, it is instead recommended to input a low value with \texttt{--f}.
\item Finally, we require that the number $\mathrm{max}(R-1)$ computed from the second half of the chains is below 0.4 for all parameters: this is the same condition as for stopping the covariance matrix update. Thus the superupdate mechanism will only stop its activity in the final ``update sequence'', during which a large number of truly Markovian steps (generated with a constant proposal density) can be accumulated. 
\end{itemize}

\noindent {\it Non-MPI user friendliness.} We implemented the superupdate mechanism in \texttt{MontePython} with the same coding principles as for the update mechanism. Thus it can also be used with or without MPI, thanks to the fact that the communication between chains works through files rather than MPI commands. In the running directory, a file  \texttt{jumping\_factors.txt} stores the sequence of all jumping factors that have been used, while the file \texttt{jumping\_factor.txt} only contains the final one, that can be used as an input value in the next run. When chains are restarted in the same directory using the \texttt{--restart} command, this will be done automatically.\\

\noindent {\it Keeping only Markovian steps in the final results.}  When \texttt{--superupdate} is activated, the code still writes some comment lines in the chains at the beginning of each new ``update cycle'', with information on the current value of $\mathrm{max}(R-1)$, $j$ and $a.r.$. Additionally, it writes an extra line of comments when the jumping factor updating stops. When analyzing the chains in \texttt{info} mode, and unless the user passes the option \texttt{--keep-non-markovian}, all the lines before will be discarded and the final numbers and plots will be based on purely Markovian chains.\\

\noindent {\it Alternative implementation for single chain runs.} The superupdate mechanism in principle requires multiple chains, since it uses convergence tests based on the Gelman-Rubin statistic. For single chain runs the code will split the chain into three separate chains in order to compute the Gelman-Rubin statistic, a practice that may be less reliable than running multiple chains. However, \texttt{MontePython} also has an alternative to superupdate that was previously implemented for single chain runs by~\cite{Schroer}. This other mechanism is activated by the flag  \texttt{--adaptive} instead of \texttt{--superupdate}. It does not use the Gelman-Rubin statistic, and it is slightly closer to the original Adaptive Metropolis algorithm of \cite{haario2001}, with an update of the covariance matrix at each single step.

\section{Fisher matrix}
\label{sec:fisher}

The well-known Fisher matrix is built from the second derivatives of the effective $\chi^2$ with respect to the model parameters computed at a minimum of the $\chi^2$, i.e. at the maximum of the likelihood (see e.g. \cite{Coe:2009xf}):
\begin{align}
\mathbf{F}_{ij} = \dfrac{1}{2} \dfrac{\partial^2 \chi^2}{\partial p_i \partial p_j} = - \dfrac{\partial^2 \ln \Lagr}{\partial p_i \partial p_j} \ .
\label{eq:Fisher}
\end{align}
By definition of the maximum likelihood point, the Fisher matrix must be positive definite and invertible. Its inverse is the covariance matrix of the Gaussian approximation to the likelihood near the best-fit point. If the matrix of second derivatives is not computed at that point, it may not be invertible. 

\subsection{Motivations for Fisher matrix computation}

We implemented in \texttt{MontePython v3.0.0} a calculation of the Fisher matrix directly from the likelihood and from eq.~(\ref{eq:Fisher}), using a finite difference method that we will detail in the next section. The motivation behind this calculation is twofold:\\

\noindent{\it Boosting MCMC runs.} The inverse Fisher matrix can be used as the input covariance matrix of an MCMC run (e.g. Metropolis-Hastings). In that case we don't need a high-accuracy calculation of this matrix, because any approximate result will likely be a good enough guess, that the Metropolis-Hastings ``update'' mechanism will quickly improve anyway. This method leads to a very significant speed up in most cases, since one rarely starts an MCMC run with already at disposal a very good covariance matrix including all pairs of parameters. Still, the method can work only if the code finds an invertible Fisher matrix in the first place, and this is only guaranteed at the exact maximum likelihood point. This condition can be easy or difficult to achieve depending on the type of run:
\begin{itemize}
\item For parameter forecasts with mock data, we usually use the fiducial spectra in the role of the observed spectra, without generating a random realisation (see e.g. \cite{Perotto:2006rj} for comments on this methodology). Thus the maximum likelihood exactly coincides with the fiducial model, known in advance by the user. Then the new Fisher method works particularly well. 
\item For parameter extraction from real data, we know at most an approximation to  the best fit point. Then, one may hope that if the distance between the true and approximate best fit points is small compared to the steps used in the finite difference method, the approximate Fisher matrix computed in the latter point will still be positive definite and invertible. We found however that this does not happen very often, so the possibility to use this new method for real data remains somewhat random. To increase chances, we incorporated in  \texttt{MontePython} a few minimum-finding algorithms taken from the \texttt{optimize} python library\footnote{If the user runs \texttt{MontePython} with the command flag \texttt{--minimize}, before using any engine (Metropolis-Hastings, Fisher, etc.), the code will re-evaluate the central starting point using a $\chi^2$ minimization algorithm. This call is done in the routine \texttt{get\_minimum()} of the module \texttt{montepython/sampler.py}. After loading some approximation for the best-fit point and for the iteration step size, this routine calls the python function \texttt{numpy.optimize.minimize()}, which accepts several values of the input parameter \texttt{method}, corresponding to different minimization algorithms. By default we did set \texttt{method=`SLSQP`}, which calls the Sequential Least SQuares Programming algorithm, but the user is free to edit the module and change one line to try different methods. The algorithm stops when the $\chi^2$ seems to be converged up to the tolerance passed through the \texttt{MontePython} input flag \texttt{--minimize-tol} (default $10^{-5}$), but there is no guarantee that the algorithm leads to the true minimum.}, that may at least help to get closer to the true best-fit point before trying the Fisher matrix computation. However, our tests show that, for the moment, the implemented minimization algorithms are not very robust, especially in the presence of many nuisance parameters.
\end{itemize}In summary, the new Fisher calculation definitely improves all MCMC forecasts (as shown in section \ref{sec:performance}), and it may also improve MCMC runs with real data (see a few examples in section \ref{sec:performance_current_data}) unless one ends up with likelihood shapes that happen to be too complicated for finding the minimum and/or running the Fisher algorithm. \\

\noindent {\it Replacing MCMC runs.} When one knows that the posterior of a given run should be nearly Gaussian, or when one is not interested in the details of the posterior (e.g. non-trivial parameter correlations with some skewness, kurtosis, banana-shape, etc.), it is tempting to replace whole MCMC parameter extraction runs by simple Fisher matrix computations. Then the inverse Fisher matrix will give some approximate one-dimensional confidence regions and two-dimensional elliptic contours. 
This is particularly straighforward for sensitivity forecasts since, in that case, the maximum likelihood point is known in advance. It can also be envisaged for real data if the maximum likelihood point is known up to good approximation, for instance, after a run with the new \texttt{--minimize} option of \texttt{MontePython} (whose success is not guaranteed).

In the cosmology literature, a vast majority of parameter forecasts are based on Fisher matrix calculations. These are usually performed by specific codes, using the fact that after a few steps of analytic calculations, $F_{ij}$ can be re-expressed as a function of the derivative of the observable quantities with respect to the parameters (e.g. $\partial C_\ell/\partial p_i$ or $\partial P(k)/\partial p_i$). Instead, the Fisher matrix computation performed by \texttt{MontePython} is a direct likelihood-based evaluation, since we compute $\partial \ln {\cal L} / \partial p_i$. The two approaches are mathematically equivalent, but the latter may offer some practical avantages. Indeed, the quantity which is primarily build to model a given experiment is the likelihood. Skipping the analytical steps leading to derivatives like $\partial C_\ell/\partial p_i$ or $\partial P(k)/\partial p_i$ sometimes avoids complicated expressions, the need to introduce approximations, and further risks to make an error.

In both approaches, one has to compute some numerical derivatives with a given step size. For a purely Gaussian likelihood, the step size should be irrelevant, provided that it is not so small that numerical errors (from the Boltzmann code or from the likelihood code) come to dominate. Very often, Boltzmann codes are optimized in order to give an accuracy on the $\chi^2_\mathrm{eff}$ of the most constraining experiments (typically, nowadays, the Planck experiment) of the order of $\delta \chi^2 \sim {\cal O}(10^{-1})$, simply because achieving better precision would not change the results on confidence intervals, and would thus be a waste of computing time. Therefore, it is dangerous to use steps $\Delta p_i$ such that
\begin{equation}
\Delta \chi^2 \equiv \frac{1}{2} \left( \ln {\cal L}(p_i^\mathrm{bestfit}+\Delta p_i) -  {\cal L}(p_i^\mathrm{bestfit}) \right) 
\end{equation}
is significantly smaller than 0.1.
This provides roughly a lower bound on the $\Delta p_i$'s. The question of the upper bound is more delicate, and especially important when the likelihood is not Gaussian: different choices can then return significantly different Fisher matrices and confidence limits. The community is split between different approaches on this issue. One school suggests to take the smallest possible steps until numerical noise comes into play, in order to be as close as possible to the mathematical definition of the second derivatives. Another school prefers to choose steps such that
$\Delta \chi^2 \simeq 1$ (resp. 4)
if the final goal is to deliver predictions for 68\% (resp. 95\%) confidence limits on the parameters, since in that case the Fisher matrix gives a Gaussian approximation of the likelihood valid precisely in the region that is relevant for the parameter bounds
(see e.g. \cite{Perotto:2006rj}). Many Fisher codes do not even target any particular given order of magnitude for $\Delta \chi^2$, and choose the steps $\Delta p_i$ arbitrarily.\\ 
In \texttt{MontePython v3.0.0}, we approach this problem by letting the user choose a target value for $\Delta \! \ln \! \Lagr = 2\Delta \chi^2$. By default, the code will first try to get an invertible Fisher matrix with $\Delta \! \ln \! \Lagr \sim 0.1$, and will iteratively increase this value in case the result is non-invertible, as explained in the next section. However, the user can choose the first value of $\Delta \! \ln \! \Lagr$, and may for instance set it to 0.5 or 2 (with the flag \texttt{--fisher-delta}).

\subsection{Iterative strategy for the fisher matrix computation}

The Fisher matrix calculation is a ``method'', like e.g. Metropolis-Hastings, Nested sampling, Importance Sampling, minimization, etc.
It is activated by launching \texttt{MontePython v3.0.0} in \texttt{run} mode with the command flag:
\begin{itemize}
\item \texttt{--method Fisher} : calculate Fisher matrix
\end{itemize}
The calculation takes place around parameter values specified by the first entries of each list in the input file:
\begin{center}\texttt{data.parameters['P1']      = [p1, .., .., .., .., ..]}  , \end{center} 
unless another best-fit model is passed with the command line \texttt{-b path/to/file.bestfit}.
The user may control the step size for the finite difference derivatives with two parameters:
\begin{itemize}
	\item \texttt{--fisher-delta} $D$: target $\Delta \! \ln \! \Lagr$ value for finding the steps $\Delta p_i$ [default: $D=0.1$]
	\item \texttt{--fisher-tol} $T$: tolerance for $\Delta \! \ln \! \Lagr$ (note: decreasing slows down computation) [default: $T=0.05$]	
\end{itemize}	
Then the code finds the step size for each parameter matching the target $\Delta \chi^2=D \pm T$ by bisection. Sometimes the bisection struggles to converge (e.g. for non-Gaussian likelihoods, or if the calculation is not centered on the maximum of the likelihood). In this case, after 10 attempts, it gradually increases the tolerance $T$ at each step until convergence is obtained. However, in such an event, it may be preferable to adjust the input parameters instead (e.g. target $D$ or best fit parameter values).

Once the step sizes have been obtained, the code computes all the elements of the Fisher matrix. If the result is a non-invertible matrix (due to the non-Gaussianity of the likelihood or to a bad guess for the maximum likelihood), the code enters into a stage of iterations over the target value  of $\Delta \chi^2$, which is steadily increased until the matrix becomes invertible, following the sequence
$D, 2D, 3D, ... ND$. The maximum number of iterations can be controlled with
\begin{itemize}
	\item \texttt{--fisher-step-it} $N$: number of step iterations attempted [default: $N=10$]
\end{itemize}
If the matrix inversion still fails after the maximum number of iterations, the code stops and returns an explicit error message.

Whenever the code finds an invertible Fisher matrix, it stores both the Fisher matrix and its inverse in distinct files with the extension \texttt{.mat}. The inverse Fisher matrix file matches the usual format of any covariance matrix that the Metropolis-Hastings algorithm would take in input for the density proposal. Thus it can immediately be used in an MCMC run with the input flag  \texttt{-c path/to/file.covmat}.

\subsection{Dealing with prior boundaries}

In the last section we mentioned that the code finds the step sizes $\Delta p_i$ used in numerical derivatives with a bisection algorithm.
The bisection starts with a first tentative step size given by the input $\sigma_i^\mathrm{input}$ value for a each parameter, as given by the input file or by the input covariance matrix specified by the input flag \texttt{-c path/to/file.covmat} (the second always has priority). In cases where a $p_i^\mathrm{bestfit} \pm \sigma_i^\mathrm{input}$ value exceeds the prior boundary, we change the initial step $\sigma_i^\mathrm{initial}$ according to the following criteria:\\
\\
\textbf{Case 0}: When there are no boundaries, or the difference between the boundary and the center is greater than $\sigma^\mathrm{input}_i$, the initial step is given by $\sigma^\mathrm{input}_i$:\\
$$\sigma^\mathrm{input}_i < B_{i-} ~~\textrm{and}~~ \sigma^\mathrm{input}_i < B_{i+} \quad \Rightarrow \quad \sigma_i^\mathrm{initial} = \sigma_i^\mathrm{input}~,$$
where the lower boundary distance is $B_{i-} = p_i^\mathrm{bestfit} - p_i^\mathrm{lower-boundary}$ and the upper one is $B_{i+} = p_i^\mathrm{upper-boundary} - p_i^\mathrm{bestfit}$ .\\
\\
\textbf{Case 1}: When one or both of the boundary distances is smaller than $\sigma_i^\mathrm{input}$, but both are still larger than a tenth of $\sigma_i^\mathrm{input}$, we set the initial step to the smaller of the two boundary distances:\\
$$0.1 \sigma_i^\mathrm{input} < B_{i-/+} < \sigma_i^\mathrm{input} \quad \Rightarrow \quad \sigma_i^\mathrm{initial} = {\rm min}(B_{i-},B_{i+})~.$$
\\
\textbf{Case 2}: When one or both of the boundary distances is smaller than a tenth of $\sigma_{input}$, we instead assume the likelihood is symmetric around the best-fit point, and we only compute steps in one direction (the one in which the distance to the boundary is the greatest), while mirroring the likelihood values to the other direction:\\
$$B_{i-/+} \leq 0.1\sigma_i^\mathrm{input}  \quad \Rightarrow \quad \sigma_i^\mathrm{initial} = {\rm min}({\rm max}(B_{i-},B_{i+}), \sigma_i^\mathrm{input})~.$$
\\
Once the steps have been settled in that way, the diagonal elements of the Fisher matrix are given by the numerical derivatives
\begin{align}
%\dfrac{\partial^2 \ln \Lagr}{\partial p_i^2} \approx \dfrac{\ln \Lagr(p_{i,+\Delta}) - 2 \ln \Lagr(p_i) + \ln \Lagr(p_{i,-\Delta}) }{\Delta p_{i}^2}.
\mathbf{F}_{ii} = \dfrac{\partial^2 \ln \Lagr}{\partial p_i^2} \approx \dfrac{\ln \Lagr(p_{i}{+\Delta p_i}) - 2 \ln \Lagr(p_i) + \ln \Lagr(p_{i}{-\Delta p_i}) }{\Delta p_{i}^2} 
\label{eq:fisher_diag}
\end{align}
and the the off-diagonal ones by
\begin{align}
\mathbf{F}_{ij} = \dfrac{\partial^2 \ln \Lagr}{\partial p_i \partial p_j} \approx \dfrac{\ln \Lagr(p_{i}{+\Delta p_i},p_{j}{+\Delta p_j}) - \ln \Lagr(p_{i}{+\Delta p_i}, p_{j}{-\Delta p_j}) - \ln \Lagr(p_{i}{-\Delta p_i},p_{j}{+\Delta p_j}) + \ln \Lagr(p_{i}{-\Delta p_i},p_{j}{-\Delta p_j})}{4 \Delta p_i \Delta p_j} \ .
\label{eq:fisher_offdiag}
\end{align}
\\
\textbf{Asymmetric steps.} In cases 0 and 1, the code always uses symmetric steps, and in case 2 it postulates a symmetry of the likelihood. In some situations the user may find it beneficial to use instead some asymmetric steps to compute the Fisher matrix.
This can be activated with the input flag:
\begin{itemize}
	\item \texttt{--fisher-asymmetric} : allow for asymmetric steps (note: slows down computation) [default: False]
\end{itemize}
Then the ``case 1'' and ``case 2'' rules are replaced with some evaluations of the likelihood at $p_i+\Delta p_{+i}$ and
$p_i-\Delta p_{-i}$, with $\Delta p_{+i} = \mathrm{min}(\sigma_i^\mathrm{input}, B_{i+})$ and $\Delta p_{-i} = \mathrm{min}(\sigma_i^\mathrm{input}, B_{i-})$. In that case the diagonal terms of the Fisher matrix are given by
\begin{align}
\dfrac{\partial^2 \ln \Lagr}{\partial p_i^2} \approx 2 \dfrac{\left(\dfrac{\Delta p_{-i}}{\Delta p_{+i}}\right) \ln \Lagr(p_{i}{+\Delta p_{+i}}) - \left(\dfrac{\Delta p_{-i}}{\Delta p_{+i}} + 1\right) \ln \Lagr(p_i) + \ln \Lagr(p_{i}{-\Delta p_{-i}}) }{\Delta p_{-i} \Delta p_{+i} + \Delta p_{-i}^2} ~,
\end{align}
and the off-diagonal ones by 
\begin{align}
\dfrac{\partial^2 \ln \Lagr}{\partial p_i \partial p_j} &\approx \left(\frac{\Delta p_{-j}^2}{\Delta p_{+j}} + \Delta p_{-j} \right)^{-1} \left(\frac{\Delta p_{-i}^2}{\Delta p_{+i}} + \Delta p_{-i} \right)^{-1} \nonumber\\
&\times \Bigg\{\left(\frac{\Delta p_{-j}}{\Delta p_{+j}}\right)^2 \left[\left(\frac{\Delta p_{-i}}{\Delta p_{+i}}\right)^2 \ln \Lagr(p_{i}{+\Delta p_{+i}},p_{j}{+\Delta p_{+j}}) - \ln \Lagr(p_{i}{-\Delta p_{-i}},p_{j}{+\Delta p_{+j}})\right] \nonumber\\
&- \left(\frac{\Delta p_{-i}}{\Delta p_{+i}}\right)^2 \ln \Lagr(p_{i}{+\Delta p_{+i}}, p_{j}{-\Delta p_{-j}}) + \ln \Lagr(p_{i}{-\Delta p_{-i}},p_{j}{-\Delta p_{-j}}) \\
&+ \left[\left(\frac{\Delta p_{-j}}{\Delta p_{+j}}\right)^2 - 1\right] \left(\ln \Lagr(p_{i}{-\Delta p_{-i}},p_{j}) - \left(\frac{\Delta p_{-i}}{\Delta p_{+i}}\right)^2 \ln \Lagr(p_{i}{+\Delta p_{+i}},p_{j}) + \left[\left(\frac{\Delta p_{-i}}{\Delta p_{+i}}\right)^2 - 1\right] \ln \Lagr(p_{i},p_{j})\right) \nonumber\\
&+ \left[\left(\frac{\Delta p_{-i}}{\Delta p_{+i}}\right)^2 - 1\right] \left(\ln \Lagr(p_{i}, p_{j}{-\Delta p_{-j}}) - \left(\frac{\Delta p_{-j}}{\Delta p_{+j}}\right)^2 \ln \Lagr(p_{i}, p_{j}{+\Delta p_{+j}}) \right) \Bigg\}~. \nonumber
%\label{eq:fisher_offdiag_asym}
\end{align}

\subsection{Efficient treatment of nuisance parameters}

Not counting the few intermediate steps necessary for the automatic determination of step sizes (which is typically around $2-4$~evaluations per parameter), the calculation of one Fisher matrix requires a number of likelihood evaluations equal to
\begin{align}
N_{\rm evaluations} = 1 + 2 N_{\rm params} + 4 \sum_{n=1}^{N_{\rm params}-1}n \ ,
\end{align}
i..e. one in the best-fit point, two for each diagonal element and four for each off-diagonal element, where $N_{\rm params} = N_{\rm cosmo} + N_{\rm nuisance}$ is the total number of parameters, $N_{\rm cosmo}$ is the number of cosmological parameters and $N_{\rm nuisance}$ is the number of nuisance parameters. For a typical Planck run we have 6 cosmological parameters and 26 nuisance parameters, resulting in 2049 likelihood evaluations when the target $\Delta \chi^2$ is not iterated on.

The fact that varying only nuisance parameters does not require a call to the Boltzmann solver allows us to considerably optimize the computation. In \texttt{MontePython}, the routine calling the likelihoods always keeps a memory of the previous step (model parameters and cosmological observables). Therefore, if the likelihood is evaluated at a new point such that only nuisance parameters have changed, \texttt{MontePython} knows that the Boltzmann code  should not be called again. To optimize the Fisher matrix calculation, we just need to arrange the $N_{\rm evaluations}$ calls of the likelihood in a particular order minimizing the number of calls to the Boltzmann solver.

We loop over the parameters starting from the cosmological ones and ending with the nuisance ones. 

For each parameter $p_i$, we first perform all the calculations involving the value $(p_{i}{-\Delta p_{i}}$, i.e.
$\Lagr(p_{i}{-\Delta p_{i}})$ for the diagonal element and
$\Lagr(p_{i}{-\Delta p_{i}},p_{j}{\pm\Delta p_{j}})$ (for each $j>i$) for the non-diagonal elements.
Then we perform all the calculations involving the value $(p_{i}{+\Delta p_{i}}$, i.e.
$\Lagr(p_{i}{+\Delta p_{i}})$ for the diagonal element and
$\Lagr(p_{i}{+\Delta p_{i}},p_{j}{\pm\Delta p_{j}})$ (for each $j>i$) for the non-diagonal elements.
In that way, the number of calls to the Boltzmann solver is drastically reduced to
\begin{align}
N_{\rm calls} = 1 + 4 N_{\rm cosmo} + 4 \sum_{n=1}^{N_{\rm cosmo}-1}n \ ,
\end{align}
corresponding to $N_{\rm calls} = 85$ for a typical Planck run (again without step iteration). This vastly reduces the computational time\footnote{Note that if the number of operation was not reordered in such a special way, we would still get some gain, but the number of calls would still be as large as
\begin{align*}
N_{\rm calls} = 1 + 2 N_{\rm cosmo} + 4 \left( \sum_{n=1}^{N_{\rm params}-1}n 
-  \sum_{m=1}^{N_{\rm nuisance}-1}m ~.
\right)
\end{align*}
This corresponds to $N_{\rm calls}=697$ for a typical Planck run, without step iteration.
}.

\subsection{Plotting likelihood contours from inverse Fisher matrix}

\begin{figure}[htb]
	\centering
	\includegraphics[width=0.75\linewidth]{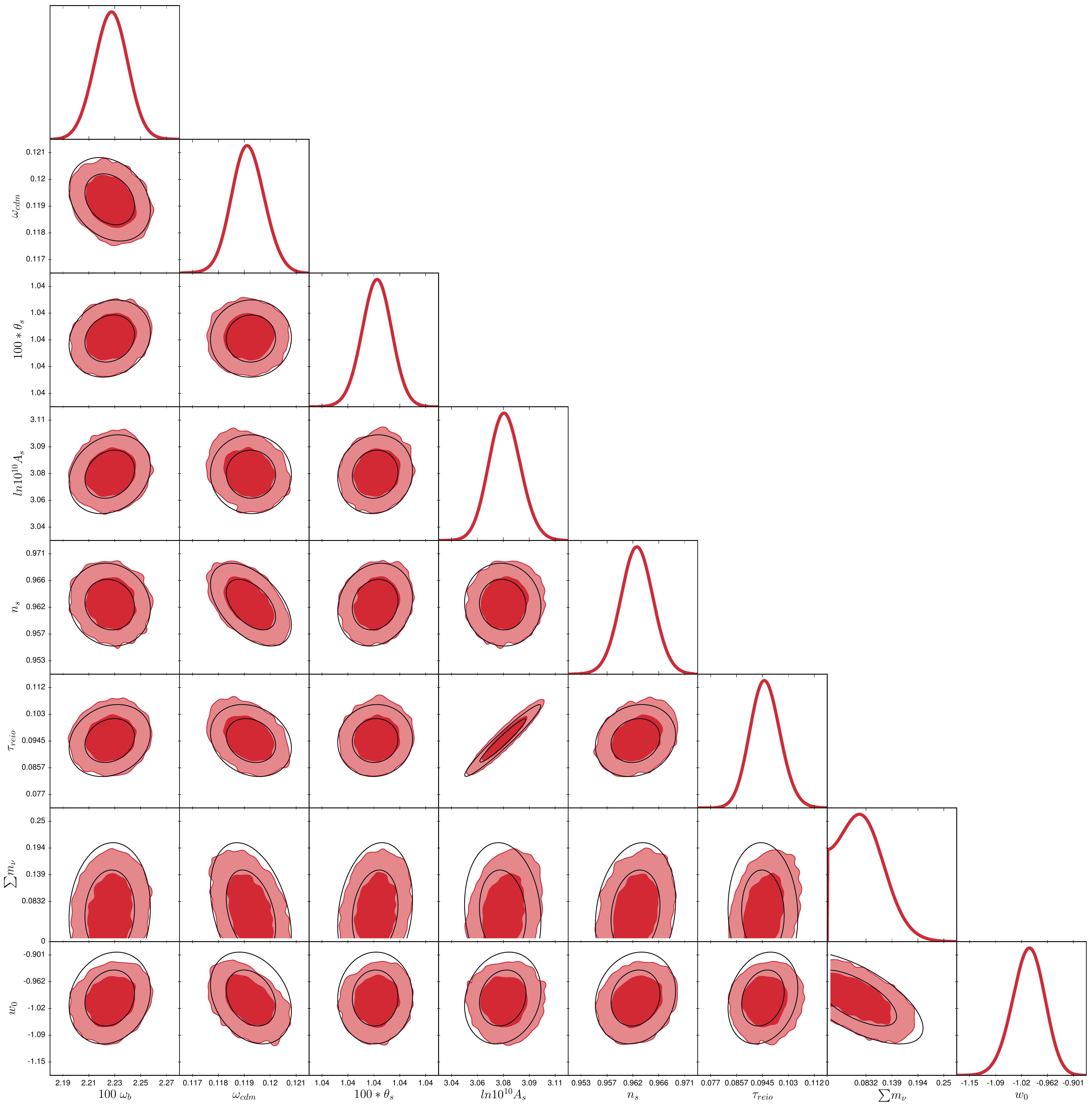}
	\caption{Confidence ellipses inferred from the inverse Fisher matrix (black lines) plotted on top of the 2d marginalized posterior distribution of a Metropolis-Hastings forecast. We fitted mock BAO data from the DESI survey combined with mock Planck data, assuming an 8 parameter cosmological model ($\nu$wCDM).}
	\label{fig:fisher_triangle}
\end{figure}

Having the Fisher matrix and its inverse, the user can easily write a small scripts to plot the ellipses corresponding to two-dimensional confidence levels in parameter space. This is not possible with \texttt{MontePython}, which can only do plots when some MCMC chains are present.

However, the \texttt{MontePython} plotting tools are useful for comparing the results of MCMC runs with the Gaussian posterior approximation given by the inverse Fisher matrix. For that purpose, one can analyse \texttt{MontePython} results with the usual \texttt{info} mode,
adding just one input flag: \texttt{--plot-fisher}. Then the code will check whether a Fisher matrix has been computed and stored in the same directory as the chains that the user is trying to plot. If this is the case, the Fisher ellipses are drawn on top of the MCMC contours, like in figure~\ref{fig:fisher_triangle}.

This figure shows a sensitivity forecast based on mock BAO data from the DESI survey combined with Planck data, for a cosmological model with massive neutrinos and dynamical dark energy ($\nu$wCDM, 8 free parameters). The Fisher matrix was actually computed before launching the chains, and its inverse was used as an input covariance matrix. The final MCMC contours prove that in this case, the Fisher approximation is excellent. This inverse Fisher matrix does not only provide a good proposal density for MCMC runs, it also gives excellent estimates of parameter bounds, and it could be substituted to the whole MCMC results.

One could be in a situation in which a Fisher matrix is first computed around a guess for the best-fit point, and then used to launch MCMC chains that will be centered on the true best-fit point (in the case of a Gaussian posterior). In the comparison plot, one may find that the Fisher ellipses have the right shape, but are offset with respect to the true best-fit point. In order to get a nicer plot, the \texttt{MontePython} user can use the input flag \texttt{--center-fisher}. This will automatically center the Fisher ellipses on the maximum likelihood point extracted from the MCMC chains, instead of using the central values read in the \texttt{log.param} file, even if the Fisher matrix was actually computed in that point.

\section{Illustration of performance}
\label{sec:performance}

In order to illustrate the performance of \texttt{superupdate} and the impact of using an inverse Fisher matrix as input covariance matrix, we have chosen a few data sets and cosmological models, and performed some fits with or without these different options. The comparison is especially interesting in the most difficult situations: large number of free parameters, small prior knowledge (i.e. poor guess for the input covariance matrix), etc..

\subsection{Forecasts with a small prior knowledge}

We first run some MCMC forecasts for the combination of mock BAO data from the DESI\footnote{\url{http://desi.lbl.gov}} survey and Planck\footnote{\url{http://sci.esa.int/planck/}} data.
We use the mock DESI likelihood called \texttt{fake\_desi\_vol} (documented in~\ref{sec:likelihoods}). For a forecast, we don't need to use real Planck data. We use instead a likelihood which simulates roughly the approximately of the Planck satellite, but uses some synthetic data corresponding to the Planck best-fit model. This likelihood is called \texttt{fake\_planck\_realistic} and is also documented in~\ref{sec:likelihoods}.

\begin{table}[thb]
\centering
\begin{tabular}{ c | c | c | c | c }
	\multicolumn{5}{ c }{Mock data: fake\_planck\_realistic, fake\_desi\_vol (see \ref{sec:likelihoods})} \\[5pt] %\hline
	\multicolumn{5}{ c }{Running time: 12 hours} \\ %\hline
	model & \# param. &  $R-1$: update &  $R-1$: superupdate &  $R-1$: superupdate + Fisher \\ \hline
	$\Lambda$CDM & 6 & 0.030 & 0.015 & 0.013 \\ \hline
	+ $\sum m_{\nu}$ + $w_0$ & 8 & 0.036 & 0.022 & 0.018 \\ \hline
	+ $N_{\rm eff}$ + running & 10 & not converged & not converged & 0.040 \\ \hline
	+ $\Omega_k$ & 11 & not converged & not converged & 0.048 \\ \hline
	+ $w_a$ & 12 & not converged & not converged & 0.088 \\ %\hline
	\multicolumn{5}{ c }{} \\[-5pt]
	\multicolumn{5}{ c }{Running time: 48 hours} \\ %\hline
	$\Lambda$CDM & 6 & 0.0035 & 0.0029 & 0.0019 \\ \hline
	$\nu w_0$CDM + $N_{\rm eff}$ + running & 10 & 0.014 & 0.0054 & 0.0038 \\ %\hline
\end{tabular}
\caption{For mock data and several cosmological models, comparison of three sampling options, using the Gelman-Rubin convergence criterium. See text for details.
\label{tab:forecast_conv}}
\end{table}

We fit these datasets with the minimal 6-parameter $\Lambda$CDM model and with several extended models featuring up to 12 free parameters. These extensions are listed in the first column of Table~\ref{tab:forecast_conv} and include massive neutrinos, dynamical dark energy with a constant equation of state, extra relativistic degrees of freedom, a running of the primordial spectrum index, spatial curvature, and finally dynamical dark energy with a CPL parametrisation \citep{Chevallier:2000qy,Linder:2002et}.

We run \texttt{MontePython} in these different cases with the Metropolis-Hastings algorithm and three different methods:
\begin{itemize}
	\item \texttt{update} [\texttt{--update}]: periodical update of the covariance matrix,
	\item \texttt{superupdate} [\texttt{--superupdate}]: additional adaptation of the jumping factor, 
	\item \texttt{superupdate} + Fisher [first \texttt{--method Fisher}; then \texttt{--superupdate}]: same but starting from the inverse Fisher matrix computed by \texttt{MontePython}.
\end{itemize}
For the \texttt{update} and \texttt{superupdate} runs, the proposal density is initialised as the ``Planck 2015 covariance matrix'', i.e. as the covariance matrix publicly distributed with the \texttt{MontePython} package, derived from the analysis of a well-converged run based on the Planck 2015 likelihoods and assuming the 6-parameter $\Lambda$CDM model. Also, in these two runs, the jumping factor is initially set to 2.4 (thus it remains equal to this value with the \texttt{update} method). For the Fisher matrix calculation, we pass to the code the exact best-fit model used to generate the mock data.

For each model and method, we launch the code with 8 chains, where each chain is running on 6 cores, using a total of 48 cores. After either 12 or 48 hours, we compute the worse \cite{Gelman:1992zz} convergence criterium ($R-1$) over all parameters, removing the initial 10-20\% of each chain (depending on the duration of the burn-in phase, but always the same for a given combination of models and experiments).

The difficulty of these runs reside in the poor guess for the input covariance matrix. In the six parameter runs, the input covariance matrix is derived from Planck data alone, while the DESI BAO data is very constraining. This means the proposal density is much too wide initially, and needs to shrink to the small region allowed by DESI data. When adding extra parameters, the situation is even worse. For the extra parameters, the code does not rely on the input covariance matrix, but on the standard deviations written in the input file (for which we plug the Planck error bars). Therefore, the proposal density needs to learn both the correct order of magnitude for the jumps in these new directions, and the parameter correlations involving the extra parameters.  

\begin{figure}[htb!]
	\centering
	\includegraphics[width=1.0\linewidth]{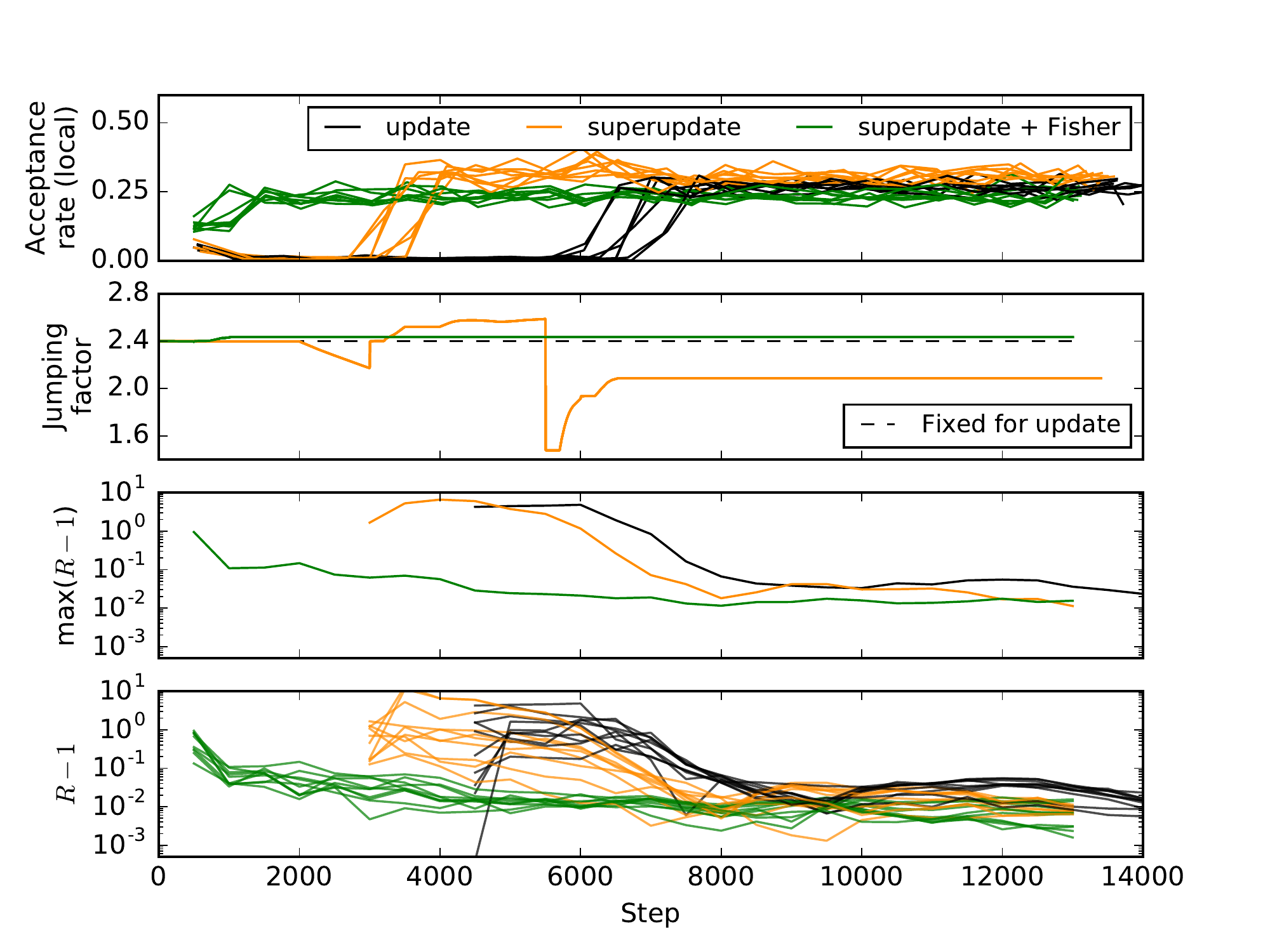}
	\caption{Evolution of the acceptance rate, jumping factor and convergence estimators for the last run from Table 1 (10 parameter model, mock Planck+DESI, 48h) using $U=50$, $S\!U=20$, FPM$\,=10$, and thus $N_\mathrm{update}=500$. The jumping factor information is updated at each step, and the information on $(R-1)$ every $N_\mathrm{update}$ step (computed over the last 50\% of the chain). This information can always be extracted from the code output. Instead, the local acceptance rate of the upper pannel was computed by post-processing the chains for the purpose of this plot, and was defined by averaging over about 500 steps (so this is slightly different from the quantity $\overbar{a.r.}$ used by the superupdate algorithms, which is only averaged over $S\!U\times\,$FPM=200 steps). The ``Fisher'' run essentially catches the right covariance matrix, jumping factor and acceptance rate from the beginning. The ``superupdate'' run reduces its jumping factor in order to quickly accumulate many points and  get a good covariance matrix estimate; once this is done, it increases the jumping factor to avoid a too big acceptance rate. Finally, the ``update'' run needs about 3500 more steps before entering into an efficient sampling regime with a good covariance matrix and acceptance rate, which corresponds to about 12 hours on our 48 cores: thus we can say that in this particular example, ``superupdate'' saved about 600~core-hours for a single run.}
	\label{10params48h}
\end{figure}

We find that, for the simplest models (6 and 8 parameters), all three methods successfully obtain at least a convergence of $R-1 = 0.03$, although \texttt{superupdate} and \texttt{superupdate} + Fisher perform better, obtaining an $R-1$ up to a factor 2 smaller.
For the more complicated models (10, 11 and 12 parameters), starting from a Fisher matrix and using \texttt{superupdate} makes a big difference, as \textit{only} the runs starting from a Fisher matrix managed to obtain any level of convergence, when limiting ourselves to only 12 hours of runtime. However, if we allow for longer runtime (48 hours) the \texttt{update} and \texttt{superupdate} methods also manage to converge, thanks to periodic updates of the covariance matrix, with the \texttt{superupdate} and \texttt{superupdate} + Fisher runs showing a factor of 2.6 to 3.7 better convergence than \texttt{update} alone. Figure~\ref{10params48h} explicitely shows why the jumping factor adaptation and the Fisher matrix calculation result in a very significant speed up for the convergence of this run.

\subsection{Current data}
\label{sec:performance_current_data}
For a comparison of the efficiency of our new methods using current data, we consider only the 6-parameter $\Lambda$CDM model,
that we fit to two data sets: a small set with just Planck and BAO likelihoods, and a larger one including Large Scale Structure (LSS) likelihoods (galaxy clustering from SDSS and weak lensing from CFHTLens). More details are given in Table~\ref{tab:data_conv}.
We perform again some fits in three different ways (\texttt{update}, \texttt{superupdate}, \texttt{superupdate}+Fisher), exactly like in the previous section (i.e. starting \texttt{update} and \texttt{superupdate} from the ``Planck 2015 covariance matrix'' distributed with the code, and from a jumping factor 2.4). We use the same number of chains and cores as in the forecasts, and allow the chains to run for 12 or 48 hours.

Like in the previous section, these runs illustrate the case of starting from a bad guess for the proposal density, because the input covariance matrix takes only Planck into account and needs to shrink to the smaller region compatible with BAO and LSS data. 
There are other significant differences with respect to the runs of the previous section. First, when we use the ``small'' dataset, we have all the nuisance parameters of the Planck high-$\ell$ TT likelihood, which have strongly non-Gaussian posteriors and are correlated with each other. This means that the optimal jumping factor is significantly different from 2.4 (it is actually closer to 1.9). It also means that the Fisher matrix calculation is difficult, due to the large number of parameters, the non-Gaussianity of the likelihood with respect to some parameters, and the fact that we only have a poor approximation of the best-fit point in parameter space (we compute the Fisher matrix in the approximate best-fit extracted from the chains of an earlier run with Planck data only). With the extended data set, the code actually fails to obtain an invertible Fisher matrix with the full Planck TTTEEE + BAO + LSS data, so we had to switch to the Planck-lite TTTEEE likelihood in order to get rid of nuisance parameters.

\begin{table}[thb]
	\centering
	\begin{tabular}{ c | c | c | c | c }
		\multicolumn{5}{ c }{Planck 2015 (highl TT, low$\ell$, lensing) + BAO (MGS, 6dFGS, LOWZ, CMASS)} \\[2pt] %\hline
		\multicolumn{5}{ c }{Running time: 12 hours} \\ %\hline
		model & \# param. &  $R-1$: update &  $R-1$: superupdate &  $R-1$: superupdate + Fisher \\ \hline
		$\Lambda$CDM & 6 & 0.019 & 0.0098 & 0.029 \\ %\hline
		\multicolumn{5}{ c }{} \\
		\multicolumn{5}{ c }{Planck 2015 (highl TTTEEE lite, low$\ell$, lensing) + BAO (MGS, 6dFGS, LOWZ, CMASS)} \\%\hline
		\multicolumn{5}{ c }{+ galaxy clustering (SDSS DR7 LRG), weak lensing (CFHTLenS)} \\[2pt] %\hline

		\multicolumn{5}{ c }{Running time: 12 hours} \\ %\hline
		%model & \# param. &  $R-1$: update &  $R-1$: superupdate &  $R-1$: superupdate + Fisher \\ \hline
		$\Lambda$CDM & 6 & 0.042 & 0.032 & 0.018 \\ %\hline
		\multicolumn{5}{ c }{} \\[-5pt]
		\multicolumn{5}{ c }{Running time: 48 hours} \\ %\hline
		$\Lambda$CDM & 6 & 0.0062 & 0.0047 & 0.0038 \\ %\hline
	\end{tabular}
	\caption{For current data and the $\Lambda$CDM model, comparison of three sampling options, using the Gelman-Rubin convergence criterium. See text for details.\label{tab:data_conv}}
\end{table}

The run with the small dataset shows the impact of the automatic jumping factor update: with \texttt{superupdate}, the code rapidly adapts the jumping factor to about 1.9, while with \texttt{update} it remains stuck at 2.4, leading to a small acceptance rate. 
Table~\ref{tab:data_conv} shows  a gain in $(R-1)$ by a factor of two when using \texttt{superupdate}.
However, this run also shows that using the inverse Fisher matrix is not always a good idea with current data and many non-Gaussian parameters, because the Fisher Matrix can be such a poor approximation of the likelihood (especially in the direction of the non-Gaussian nuisance parameters) that it is actually a worse input covariance matrix than the one derived from MCMC chains for a previous Planck-only run. Therefore, the preliminary Fisher calculation degrades the performance by a factor three compared to  \texttt{superupdate} + the ``Planck 2015 covariance matrix''. 

The run with the large dataset (but with the Planck-lite TTTEEE likelihood) shows instead the same trend as the forecasts: both \texttt{superupdate} and Fisher bring significant improvement , by up to a factor two in $(R-1)$. 

These different situations bring us to the following conclusion, which match several other tests that we have performed and not included here:
\begin{itemize}
\item using \texttt{superupdate} is essentially always a good idea. The only situations in which one could consider sticking to \texttt{update} are the easiest ones, i.e. when a new run involves a dataset and a model so similar to a previous run that we already have an excellent knowledge of the covariance matrix and of the optimal jumping factor. In that case, \texttt{update} and \texttt{superupdate} are nearly equivalent, but in the most unlucky situations, $\texttt{superupdate}$ could have a transitory phase during which the jumping factor would go away from the optimal value before going back to it asymptotically, and would lose a bit in efficiency. This is normally marginal and we can safely recommend to use \texttt{superupdate} in all cases: then, depending on the ``difficulty'' of the run, the improvement will range from negligible to large.
\item when there are many non-Gaussian parameters, such as the Planck nuisance parameters, the Fisher matrix computation often fails, and even when it does not fail, the inverse Fisher matrix is often a bad approximation for the proposal density, compared to any input covariance matrix that was inferred from chains with the same nuisance parameters. For example, this means that when using the full Planck high-$\ell$ likelihoods one should use the distributed ``Planck 2015 covariance matrices'', or one's own covariance matrices from previous runs, instead of the Fisher option. In almost all other cases, we found that computing and starting from the inverse Fisher matrix is a very powerful way to speed up convergence.
\end{itemize}
\section{Summary and conclusions}
We find that using \texttt{superupdate} and an inverse Fisher matrix as input covariance matrix reduces convergence time in most cases, and makes the process of obtaining convergence significantly simpler, due to much fewer trial and error runs being necessary.

The Fisher matrix computation is very quick, so we recommend that for forecasts (where the minimum is known) with a Metropolis-Hastings algorithm are preceded by a Fisher matrix computation, when an accurate covariance matrix is not available beforehand. Likewise, \texttt{superupdate} generally performs equal to or better than \texttt{update} alone, as it optimizes the acceptance rate, and we recommend its use for \textit{all} Metropolis-Hastings runs.

We acknowledge that the calculation the Fisher matrix is not entirely robust in \texttt{MontePython v3.0.0}, since in difficult cases with many non-Gaussian parameters (such as the Planck nuisance parameters), the Fisher matrix found by the code can be non-invertible, due to numerical errors and/or a poor estimate of the best-fit parameter values. We expect that further progress can be made in the future in the minimization and Fisher algorithms. However, we have tested our new features in hundreds of runs (including the few cases detailed in section \ref{sec:performance}) and found them extremely convenient for saving CPU time.

\section*{Acknowledgements}
We wish to thank Maria Archidiacono, S\'ebastien Clesse, Benedikt Schr\"oer, Tim Sprenger and Thomas Tram for very useful input or feedback. This work was supported by the Deutsche Forschungsgemeinschaft through the graduate school ``Particle and Astroparticle Physics in the Light of the LHC'' and through the individual grant ``Cosmological probes of dark matter properties''. The runs for this paper have been using the facilities provided by the RWTH High Performance Computing cluster under the project \texttt{rwth0113}.

\appendix

\section{Extra parameterisations}
\label{sec:parametrizations}

By design, any function in \texttt{CLASS} that has been incorporated into the \texttt{CLASS} python wrapper\footnote{Note that ``incorporating'' a new \texttt{CLASS} parameter in the wrapper just consists of adding one line in \texttt{python/cclassy.pxd} with just a declaration of this parameter (e.g. \texttt{double my\_param}). The declaration must be done within the structure to which the parameter belongs. Incorporating a new \texttt{CLASS} function also boils down to declaring it in this file. New coding in the file \texttt{python/classy.pyx} is only required when one wants to create a new function specific to the wrapper itself, rather than just interfacing a \texttt{CLASS} function.} can be accessed directly from \texttt{MontePython}, without any additional coding required. However, sometimes it can be useful to define a specific parametrization within \texttt{MontePython}. This is easily done in the \texttt{python/data.py} module, where the \texttt{MontePython} input parameters can be intercepted and reshaped/renamed before being passed to \texttt{CLASS}. In the \texttt{update\_cosmo\_arguments()} function, all varying cosmological parameters are iterated through, and any additional parametrisations that are desired can be included by adding a simple \texttt{if} statement similar to existing ones. 

\texttt{MontePython} includes several reparameterisation of this type. Some of them just deal with ordinary cosmological parameters, e.g.:
\begin{itemize}
\item if the parameter \texttt{Omega\_Lambda} is used as a \texttt{MontePython} input parameter, instead of being passed to \texttt{CLASS}, it is used for defining \texttt{h} according to the rule $h=\sqrt{(\omega_\mathrm{b}+\omega_\mathrm{cdm})/(1-\Omega_\Lambda)}$.
\item if the parameter \texttt{Omega\_L} is used as a \texttt{MontePython} input parameter, instead of being passed to \texttt{CLASS}, it is used for defining \texttt{omega\_cdm} according to the rule $\omega_\mathrm{cdm}=(1-\Omega_\Lambda) h^2 - \omega_\mathrm{b}$.
\item if the parameter \texttt{ln10\^{}\{10\}A\_s} is used as a \texttt{MontePython} input parameter, instead of being passed to \texttt{CLASS}, it is used for defining \texttt{A\_s}.
\item if the parameter \texttt{exp\_m\_2\_tau\_As} ($\equiv e^{-2 \tau_\mathrm{reio}} A_s$) is used as a \texttt{MontePython} input parameter, instead of being passed to \texttt{CLASS}, it is used for defining \texttt{A\_s} (assuming that \texttt{tau\_reio} is also being used).
\end{itemize}
The code includes many other redefinitions related to isocurvature modes, neutrinos, dark energy, etc. Below we expand the discussion concerning a few of the implemented neutrino and Dark Energy reparameterisations. The user can easily extend the list of reparameterisation for her/his specific cases. 

\subsection{Neutrino hierarchy}
\texttt{MontePython} can sample the total neutrino mass, with the individual neutrino masses arranged according to the Normal Hierarchy (NH, with two less massive and one more massive neutrino) or Inverted Hierarchy (IH, with one less massive and two more massive neutrinos).

The quantities passed to \texttt{CLASS} are the individual neutrino masses, but the quantity we are interested in sampling is the sum of neutrino masses. Formally, this is done using as a varying \texttt{MontePython} parameter \texttt{M\_tot\_NH} or \texttt{M\_tot\_IH}. Then, for each sampled value of the total neutrino mass ($M_{\nu}$), the individual neutrino masses ($m_{i}$) are calculated by solving the system of equations (see e.g. \cite{Hannestad:2016fog})
\begin{align*}
&M_{\nu} = m_1 + m_2 + m_3 \ ,\\
&\Delta m^{2}_{\rm atm} = m_{3}^2 - m_{\ell}^2 \ ,\\
&\Delta m^{2}_{\rm sol} = m_{2}^2 - m_{1}^2 \ ,
\end{align*}
where $\ell$ is 1 for NH and 2 for IH, and $\Delta m^{2}_{\rm atm}$ and $\Delta m^{2}_{\rm sol}$ are the current central values of the mass splittings obtained from neutrino oscillation experiments~\citep{Esteban:2016qun} (for a more recent study see e.g.~\cite{Gariazzo:2018pei})
\begin{align*}
\textbf{NH:}&\quad \Delta m^{2}_{\rm atm} = 2.524 \times 10^{-3}\ \rm eV^2 \ ,\\
&\quad \Delta m^{2}_{\rm sol} = 7.50 \times 10^{-5}\ \rm eV^2 \ ,\\
\textbf{IH:}&\quad \Delta m^{2}_{\rm atm} = -2.514 \times 10^{-3}\ \rm eV^2 \ ,\\
&\quad \Delta m^{2}_{\rm sol} = 7.50 \times 10^{-5}\ \rm eV^2 \ .
\end{align*}
Additionally, the three parameters \texttt{N\_ur=0.00641}, \texttt{N\_ncdm=3} and \texttt{T\_ncdm='0.71611,0.71611,0.71611'} should be fixed as \texttt{cosmo\_arguments}, to reflect the fact that have three distinct standard active neutrino species and no extra relativistic degrees of freedom (unless one is studying a scenario with extra relativistic relics, in which case \texttt{N\_ur} should be varied). Note that \texttt{T\_ncdm} gives, for each species, the temperature of the neutrinos in units of photon temperature. In reality, the standard neutrinos distribution contains slightly non-thermal distribution, while by default \texttt{CLASS} will treat them as thermal species. The price to pay is to have a temperature  ratio adjusted to 0.71611 in order to get the right neutrino density in the non-relativistic regime, and a small contribution to \texttt{N\_ur} in order to get the right density in the relativistic one. This is why with three massive species we advise to take \texttt{N\_ur=0.00641} instead of \texttt{N\_ur=0}, but this correction is anyway well below the sensitivity of current experiments.

\subsection{Degenerate massive $\nu$'s and varying N$_{\rm eff}$}
In addition to arranging the mass of the neutrinos in a neutrino hierarchy, it is possible to sample the total neutrino mass for a case with three massive neutrinos with degenerate mass. Although not a realistic scenario, it is often sufficient to use three degenerate neutrinos (see e.g. \cite{DiValentino:2016foa}), speeding up computations in the Boltzmann solver.

This is done via the input parameter \texttt{M\_tot}, remembering to specify the \texttt{cosmo\_arguments} from before, but this time with only one type of neutrino species, \texttt{N\_ur=0.00641}, \texttt{N\_ncdm=1} and \texttt{T\_ncdm=0.71611}, and instead specifying the degeneracy of the neutrino species, \texttt{deg\_ncdm=3}. The total neutrino mass is then simply divided by the number of massive neutrino species and the resulting particle mass is passed to \texttt{CLASS}.

Additionally, this allows for varying the effective number of relativistic species, N$_{\rm eff}$, by using the degeneracy of neutrino species, \texttt{deg\_ncdm}, as a varying cosmological parameter instead of a fixed quantity.

For completeness, it is possible to use only one or two degenerate massive neutrinos and the rest massless, but this has been shown to be slightly inaccurate for the precision of current experiments \citep{Giusarma:2016phn}.

\subsection{Dynamical dark energy}

Many phenomenological dark energy models can be treated using the fluid sector of \texttt{CLASS}, which has several free parameters labelled as \texttt{\_fld}. By default, this sector uses the PPF parameterisation \citep{Fang:2008sn}, although real fluid equation can be restored by setting
\texttt{use\_ppf} to \texttt{'no'} in the \texttt{cosmo\_arguments}.

In principle, the dynamical dark energy equation of state parameters \texttt{w0\_fld} and \texttt{wa\_fld}, with a CPL parameterisation \citep{Chevallier:2000qy,Linder:2002et} defined through $w(a) = p_{DE} / \rho_{DE} = w_0 + w_a (1 - a / a_0)$ (where, as usual, $p$ is the pressure, $\rho$ is the density, and $a$ is the scale factor), can be passed directly to \texttt{CLASS}. However, it may be useful to sample the quantity $w_0 + w_a$ (implemented as \texttt{w0wa}) and $w_0$ (implemented \texttt{w0\_fld}), in order to restrict the parameter space of $w_0 + w_a$ to only negative values (as in e.g. \cite{Upadhye:2017hdl}).

\subsection{Sterile $\nu$ parametrization}
A final example of how a specific parameterisation can be introduced in MontePython is sterile neutrinos. In addition to degenerate massive neutrinos, one may wish to sample the sterile neutrino mass and the contribution of sterile neutrinos to N$_{\rm eff}$, while avoiding the region of parameter space where the sterile neutrino mass becomes arbitrarily large and the contribution to N$_{\rm eff}$ becomes arbitrarily small (see Fig. 32. of \cite{Ade:2015xua}).

For this we defined the effective sterile neutrino mass $m_{s,\rm eff} \equiv m_{s} \Delta N_{s}$ as a possible varying cosmological parameter (\texttt{m\_s\_eff}), that should be used along with the parameter \texttt{deg\_ncdm\_\_2} standing for the contribution of sterile neutrinos to N$_{\rm eff}$. This case is a bit more complicated than the others, as, in addition to setting \texttt{N\_ncdm=2} and \texttt{T\_ncdm='0.71611,0.71611'}, we also need to set the degeneracy of normal neutrinos \texttt{deg\_ncdm\_\_1} as a ‘phantom' varying cosmological parameter, but with the parameter \textbf{fixed to 3}. The mass of the degenerate active neutrino species can also be varied as \texttt{m\_ncdm\_\_1}. This means the output of the chains will be the mass of a single active neutrino, rather than the sum, but of course we know that in this case $M_{\nu} = \texttt{m\_ncdm\_\_1} \times \texttt{deg\_ncdm\_\_1} = 3 \,\, \texttt{m\_ncdm\_\_1}$.

The effective sterile neutrino mass is then converted to physical sterile neutrino mass within the \texttt{data.py} module, in the function \texttt{update\_cosmo\_arguments()}, by dividing with $\Delta N_s$ (assuming that this is the ncdm species number 2 and that it is Dodelson-Widrow-like, i.e with the same temperature as active neutrinos). It is finally passed to \texttt{CLASS} along with the other neutrino masses.
\section{Sampling options}
\label{sec:sampling_options}

\noindent \texttt{MontePython} has the following general sampling options

\begin{itemize}
	\item \texttt{--method} : sampling method (MH, NS, CH, IS, Der, Fisher) [default: MH]\\
	which refer respectively to Metropolis-Hastings, Nested Sampling (= MultiNest), Cosmo Hammer (= emcee), Importance Sampling, Derived (= reprocessing the chains to add columns with extra derived parameters requiring a new \texttt{CLASS} run for each model), and Fisher.
	\item \texttt{-T} : sample from the probability distribution $P^{1/T}$ instead of P [default: 1.0]
\end{itemize}

\noindent Options for Metropolis-Hastings and variants
\begin{itemize}
	\item \texttt{--method MH} : Metropolis-Hastings sampling [default: MH]
	\item \texttt{--update} : proposal distribution update frequency in number of cycles [default: 50]
	\item \texttt{--superupdate} : also adapt jumping factor. Adaptation delay in number of cycles [default: 0] (i.e. deactivated by default. Recommended: 20)
	\item \texttt{--superupdate-ar} : target local acceptance rate [default: 0.26]
	\item \texttt{--superupdate-ar-tol} : tolerance for local acceptance rate [default: 0.01]
	\item \texttt{--adaptive} : running adaptation of covariance matrix and jumping factor (note: only suitable for single chain runs) [default: 0]
	\item \texttt{--adaptive-ts} : starting step for adapting the jumping factor [default: 1000]
	\item \texttt{-f} : jumping factor [default: 2.4]
	\item \texttt{--minimize} : attempt to re-evaluate starting point using a $\chi^2$ minimization algorithm [by default uses \texttt{SLSQP} via \texttt{numpy.optiminize.minimize()}, can be changed in \texttt{sampler.py} function \texttt{get\_minimum()}]
         \item \texttt{--minimize-tol} : tolerance for minimization [default: $10^{-5}$]
\end{itemize}

\noindent Fisher matrix options
\begin{itemize}
	\item \texttt{--method Fisher} : compute a Fisher matrix [default: MH]
	\item \texttt{--fisher-asymmetric} : allow for asymmetric steps (note: slows down computation) [default: False]
	\item \texttt{--fisher-step-it} : number of step iterations attempted [default: 10]
	\item \texttt{--fisher-delta} : target $\Delta \ln \Lagr$ value for step iteration [default: 0.1]
	\item \texttt{--fisher-tol} : tolerance for $\Delta \ln \Lagr$ (note: decreasing slows down computation) [default: 0.05]
	\item \texttt{--fisher-sym-lkl} : cut-off for switching to symmetric likelihood assumption in units of $\sigma$. Relevant when parameter space boundaries are close to the central value [default: 0.1]
\end{itemize}

\texttt{MontePython} also supports sampling with \texttt{MultiNest} (\texttt{--method NS}) \citep{Feroz:2007kg, Feroz:2008xx, Feroz:2013hea} via a python wrapper \citep{Buchner:2014nha} and \texttt{emcee} \citep{2010CAMCS...5...65G,ForemanMackey:2012ig,Akeret:2012ky} via \texttt{CosmoHammer} (\texttt{--method CH}) \citep{Akeret:2012ky}. For these sampling options we refer to the official documentation of those codes.
\section{Analyze and plotting options}
\label{sec:plotting_options}

\noindent The range of plotting options and the general presentation of the plots has been significantly improved in \texttt{MontePython v3.0.0}. We should, however, point out that the user is free to use other plotting tools, if she/he prefers. In particular, the \texttt{MontePython} output is fully compatible with Antony Lewis's \texttt{GetDist}\footnote{\url{http://getdist.readthedocs.io/en/latest/}}. Note that \texttt{MontePython} writes in each output directory a file in the \texttt{.paramnames} format just for this purpose. \texttt{GetDist} has some very advanced plotting functionalities and a very nice graphical interface. However, the user will benefit from a few advantages when using the \texttt{MontePython} analyzing and plotting tools, such as: automatically evaluating the burn-in phase; automatically eliminating the non-Markovian part of the chains; and automatically reading information regarding the parameter names, ranges and scalings in the \texttt{log.param} file.

\subsection{Chain analysis}

When analyzing the chains, \texttt{MontePython} eliminates automatically the burn-in phase at the beginning of each chain, before applying additional cuts that can be customised with the options listed below. The burn-in phase of each chain is defined as: all the first points in the chains until an effective $\chi^2$ value smaller than $\chi^2_\mathrm{min}+6$ was reached for the first time. This number of 6 can be adjusted manually (it is equal to 2~\texttt{LOG\_LKL\_CUTOFF}, where \texttt{LOG\_LKL\_CUTOFF} is a parameter set in \texttt{montepython/analyze.py}, with a default value of 3). For runs in which a good estimate of the best-fit model was passed in input (with the option \texttt{-b <xxx>.bestfit}), the burn-in phase defined in this way may not exist at all.

Additionally, \texttt{MontePython} has the following options for analyzing chains (thus they should be written after the command line \texttt{python montepyhton/MontePython.py info}):

\begin{itemize}
	\item \texttt{--keep-non-markovian} : keep the non-Markovian part of the chains [default: False].
	\item \texttt{--keep-fraction} : pass a decimal fraction, e.g. 0.8 to keep the last 80 \% of the part of the chains that remain after the burn-in removal (note: redundant if non-Markovian points are discarded) [default: 1.0] 
	\item \texttt{--want-covmat} : compute a covariance matrix based on the chains (note: this will overwrite the one produced by \texttt{--update}) [default: False]
	\item \texttt{--bins} : the number of bins for computing histograms [default: 20]
	\item \texttt{-T} : raise posteriors to the power T [default: 1.0]
	\item \texttt{--silent} : do not write any standard output (useful when running on clusters) [default: False]
	\item \texttt{--minimal} : use this flag to avoid computing posteriors, confidence limits and plots. The code just analyses the chains and outputs the files containing the convergence statistics, the best-fit parameters, and possibly the covariance matrix if \texttt{--want-covmat} is on [default: False]
\end{itemize}

Updating the proposal distribution or jumping parameter means that all prior steps in the chain are no longer Markovian, i.e. that each step should not depend on any prior steps. However, by using appropriate criteria for stopping adaptation of the jumping parameter and proposal distribution, and only including all steps after this point in our final analysis, we can ensure that our process was still Markovian. This is automatically done, but can be disabled with the command \texttt{--keep-non-markovian}, especially in slowly converging cases, when the user struggles to get a good covariance matrix that would stop the updating process, and wants to see some approximate results anyway.
Although the burn-in phase is always removed, if non-Markovian steps are included the user may want to use the command \texttt{--keep-fraction <number>} in order to remove the first part of the chain.\\

\subsection{Basic plotting}

\noindent The most basic plotting features are implemented as command line options (but many of them can also be passed through an input customisation file, as we shall in \ref{sec:adv_plot}):
\begin{itemize}
	\item \texttt{--no-plot} : disable plotting [default: False]
	\item \texttt{--no-plot-2d} : only plot 1d posterior distributions [default: False]
	\item \texttt{--all} : output all individual 2D subplots and histogram files as separate files
	\item \texttt{--ext} : format and extension of the plot files (pdf, eps, png) [default: pdf]
	\item \texttt{--no-mean} : in 1D plot, do not plot the ``mean likelihood'' as dashed lines, only plot the posteriors as solid lines [default: False]
	\item \texttt{--contours-only} : line contours instead of filled contours [default: False]
	\item \texttt{--posterior-smoothing} : smoothing scheme for 1D posteriors: 0 means no smoothing, 1 means cubic interpolation, $n>1$ means fitting $\ln({\cal P})$ with a polynomial of order $n$ [default: 5]
	\item \texttt{--interpolation-smoothing} :
                        for 2D contours only, interpolation factor for getting a finer histogram before applying Gaussian smoothing and getting contours; 1 means no interpolation, increase for finer curves [default: 4]
        \item \texttt{--gaussian-smoothing} :
                        for 2D contours only, width of Gaussian smoothing applied to histogram before getting contours, in units of bin size; increase for smoother contours, decrease for more exact results [default: 0.5]
	\item \texttt{--short-title-1d} : short 1D plot titles. Remove mean and confidence limits above each 1D plots. [default: False]
	\item \texttt{--num-columns-1d} : for 1D plots, number of plots per horizontal raw; if 'None' this is set automatically (trying to approach a square plot) [default: None]
	\item \texttt{--fontsize} desired fontsize [default: 16]
        \item \texttt{--ticksize} desired ticksize [default: 14]
        \item \texttt{--line-width} set line width [default: 4]
        \item \texttt{--decimal} number of decimal places on ticks [default to 3]
        \item \texttt{--ticknumber} number of ticks on each axis [default to 3]
        \item \texttt{--legend-style} specify the style of the legend, to choose from `sides` or `top` [default: sides]
\end{itemize}
\noindent When an Inverse Fisher matrix has been computed \texttt{--method Fisher}, the Fisher ellipses can be plotted on top of MCMC contours using the plotting options:
\begin{itemize}
	\item \texttt{--plot-fisher} : plot inverse Fisher matrix contours [default: False]
	%\item \texttt{--use-fisher-it} : specify inverse Fisher matrix iteration number to plot [default: 1]
	\item \texttt{--center-fisher} : centers Fisher ellipses on the parameters extracted from the best-fit model found in the chains, instead of the central starting values found in the input file [default: False]
\end{itemize}

\subsection{More advanced plot customisation \label{sec:adv_plot}}

\noindent Further options for customizing plots can be passed through a file with extension \texttt{.plot} called with the option \texttt{--extra}. All functionalities are mentioned in the example file \texttt{plot\_files/example.plot} that the user would call with the plotting option 
\texttt{--extra plot\_files/example.plot}. Although this file is self-explanatory, we list here the main functionalities provided by the use of \texttt{.plot} files. Several options have been present since the first release of \texttt{MontePython}:
\begin{itemize}
\item on-the-fly redefinition of the chain parameters with a simple syntax. For instance, if you know that there is a parameter called \texttt{A} and one called \texttt{B}, you can in principle replace the numbers in the column  \texttt{A} by the result of any algebraic operation involving \texttt{A} alone, or \texttt{A} and \texttt{B}, or even more parameters, like e.g. \texttt{A} + 3 \texttt{A} /\texttt{B}. The file \texttt{plot\_files/example.plot} provides the following example:\\
\mbox{   }~~~~~\texttt{info.redefine = \{'omega\_cdm': '(0.01*omega\_b+omega\_cdm)/(H0/100.)**2'\} }\\
In this example, the code takes the numbers in the column \texttt{omega\_b} and first multiplies them by 0.01, knowing that in the chains, $\omega_b$ was rescaled by $100$ (this actually depends on what the user wrote in the input file). Thus, \texttt{0.01*omega\_b} is the true $\omega_\mathrm{b}$, and \texttt{(0.01*omega\_b+omega\_cdm)/(H0/100.)**2} is in fact $\Omega_\mathrm{m}$. With the above command, each value of $\omega_\mathrm{cdm}$ is replaced on-the-fly by $\Omega_\mathrm{m}$ when the chains are read. The next necessary step is to change the name of the parameter for this column from \texttt{omega\_cdm} to \texttt{Omega\_m}, which can be done by the next functionality.
\item redefinition of parameter name, for the purpose of redefinitions or making the parameter name better readable by the LaTeX routines of the plotting algorithm, e.g.\\
\mbox{   }~~~~~\texttt{ info.to\_change = \{'omega\_cdm': '\$Omega\_\textbackslash mathrm\{m\}\$' \} }\\
will replace \texttt{omega\_cdm} with \texttt{Omega\_m}.
Note that, for the purpose of getting a nice LaTeX format, \texttt{MontePython} already does several basic operations automatically, like identifying greek letters, subscripts and superscripts. Hence, at the time of producing a plot label, it would automatically convert \texttt{omega\_cdm} into \texttt{\$\textbackslash omega\_\{cdm\}\$}. The functionality \texttt{info.to\_change} is useful in order to further customise the LaTeX formatting.
\item redefine the overall rescaling factor when the one from the input file is not optimal (scaling factors are useful e.g. to get rid of powers of ten in the plot captions, for very small or large parameters). This is done with the syntax \texttt{info.new\_scales = \{'A': 100\}}.
\item specify the list of parameters to be plotted (taking into account the new names, if there were name redefinitions). This is done with the \texttt{info.to\_plot = [...]} syntax, which is very useful e.g. for getting rid of nuisance parameters in the 1D and 2D plots.
\end{itemize}
The new functionalities in \texttt{MontePython} \texttt{v3.0.0} are:
\begin{itemize}
\item parameters to control the legends: \texttt{info.plot\_legend\_1d}, \texttt{info.plot\_legend\_2d}, \texttt{info.legendnames}\\ (see \texttt{plot\_files/example.plot} for details).
\item parameters to control the colors: \texttt{info.MP\_color\_cycle}, \texttt{info.MP\_color}, \texttt{info.alphas}\\ (see \texttt{plot\_files/example.plot} for details).
\item these lines simply overwrite the value of some parameters defined previously by the code within the python class \texttt{info}. Many other such lines can be added there, for instance \texttt{info.ticknumber = 5}, etc. Thus some of the options described previously as command line options can also be passed here, as lines of python.
\item sometimes, the user would like to add some extra lines of python code in the plotting script, in order to further customise 1D or 2D plots, e.g. with vertical or horizontal lines, bands, arrows, labels, etc.. Usually, these lines are meant for only specific 1D or 2D plots. One can now achieve this by writing a few extra lines of python code in little files with a \texttt{.py} extension, which will be read and executed before finalizing the relevant plots. If they start with appropriate \texttt{if} statements, they will only be taken into account when plotting specific parameters. Some self-explanatory examples are provided together with the code in the files \texttt{plot\_files/example.plot}, \texttt{add\_h\_contour.py}, and \texttt{add\_sigma8\_Omegam\_contour.py}.
\end{itemize}

\newpage
\section{Likelihoods}
\label{sec:likelihoods}
Below is a comprehensive list of the likelihoods in the \texttt{MontePython} v3.0.0 package, as well as references to the paper(s) that should be cited when used (i.e. either where the likelihood was published and/or first used with \texttt{MontePython}). We recall that it is easy to modify these likelihoods or to create new ones. Some guidelines are given in the \texttt{MontePython} online documentation\footnote{\url{http://monte-python.readthedocs.io}}, in the section ``Existing likelihoods, and how to create new ones''.
The column LU (Last Updated) shows the version number of the last modification. 

In the column D (Dependencies), SC stands for self-contained; D means that some external data files must be downloaded; W means that we provide a wrapper to some external likelihood code that must be downloaded together with some data (as e.g. for Planck likelihoods); M means that this likelihood will automatically generate its own mock data, unless it has already been generated by a previous run. In the cases D, W, M, if you run the likelihood before downloading the required external files or before having created mock data, a self-explanatory message will tell you where to download from or what to do. 

\begin{table}[h!]
	\begin{tabular}{ l | c | c | c | c | c }
		\multicolumn{6}{ c }{Current data likelihoods} \\ %\hline
		name & description & type & LU & D &reference(s) \\ \hline \hline
		acbar & ACBAR 2017 & CMB & 1.0 & SC & \cite{Kuo:2006ya} \\ \hline
		bao & 6dFGS & BAO & 1.1 & SC & \cite{Beutler:2011hx} \\
		& BOSS DR9,  & & & & \cite{Anderson:2012sa} \\ 
		& SDSS DR7 & & & & \cite{Ross:2014qpa} \\ \hline
		bao\_known\_rs & same as bao assuming  & BAO & 1.1 & SC & \cite{Audren:2013nwa} \\
		& known sound horizon value & & & & \\ \hline
		bao\_angular & angular 2-point & BAO & 3.0 & SC & \cite{Sanchez:2010zg} \\
		& correlation function & & & & \cite{Carvalho:2015ica} \\
		& SDSS DR7: LRG & & & & \cite{Alcaniz:2016ryy} \\
		& BOSS DR10\&11: CMASS & & & & \cite{deCarvalho:2017xye} \\
		& BOSS DR12: QSO & & & & \cite{Carvalho:2017tuu} \\ \hline
		bao\_boss & 6dFGS, & BAO & 2.0 & SC & \cite{Beutler:2011hx} \\ 
		& BOSS DR10\&11: & & & & \cite{Anderson:2013zyy} \\ 
		& LOWZ, CMASS, & & & & \cite{Ross:2014qpa} \\
		& SDSS DR7: MGS & & & & \\ \hline
		bao\_boss\_aniso & BOSS DR10\&11: CMASS& BAO & 2.0 & SC & \cite{Anderson:2013zyy} \\ \hline
		bao\_boss\_aniso\_gauss\_approx & BOSS DR10\&11: CMASS & BAO& 2.0 & SC & \cite{Anderson:2013zyy} \\ \hline
		bao\_boss\_dr12 & BOSS DR12: & BAO & 3.0 & SC & \cite{Alam:2016hwk} \\
		& LOWZ \& CMASS & & & & \cite{Buen-Abad:2017gxg} \\ \hline
		bao\_fs\_boss\_dr12 & BOSS DR12: & BAO+RSD & 3.0 & SC & \cite{Alam:2016hwk} \\
		& LOWZ \& CMASS & & & & \cite{Buen-Abad:2017gxg} \\ \hline
		bao\_smallz\_2014 & 6dFGS, & BAO & 3.0 & SC & \cite{Beutler:2011hx} \\
		& SDSS DR7: MGS & & & & \cite{Ross:2014qpa} \\ \hline
		%bbn & Planck? & Prior: $\omega_b$ & & \\ \hline
		bicep & BICEP & CMB & 1.0 & SC & \cite{Chiang:2009xsa} \\ \hline
		bicep2 & BICEP2 & CMB & 2.0 & SC & \cite{Ade:2014xna} \\ \hline
		BK14 & Bicep-Keck-Planck 2014 & CMB & 3.0 & D & \cite{Array:2015xqh} \\ \hline
		BK14priors & priors for the latter & CMB & 3.0 & D & \cite{Array:2015xqh} \\ \hline
		boomerang & BOOMERanG & CMB & 1.0 & SC & \cite{Jones:2005yb} \\ \hline
		cbi & CBIpol & CMB & 1.0 & SC & \cite{Sievers:2005gj} \\ \hline
		CFHTLens & CFHTLens as $\Omega_m^\alpha \sigma_8$ prior& Weak Lens. & 2.1 & SC & \cite{Heymans:2013fya}  \\ \hline
		CFHTLens\_correlation & full CFHTLens correlation & Weak Lens. & 2.2 & SC & \cite{Heymans:2013fya}
	\end{tabular}
	\caption{Current data likelihoods (letters a-c) \label{tab:likelihoods_data}}
\end{table}

\begin{table}[p!]
	\begin{tabular}{ l | c | c | c | c | c }
		name & description & type & LU & D & reference(s) \\ \hline \hline
		clik\_wmap\_full & WMAP 7yr & CMB & 1.2 & W & \cite{Larson:2010gs} \\ 
		& (through Planck wrapper) & & & & \\ \hline 
		clik\_wmap\_lowl & WMAP 7yr: low $\ell$ & CMB & 1.2 & W & \cite{Larson:2010gs} \\ \hline
		cosmic\_clocks\_2016 & cosmic clocks & $H(z)$ &  3.0 & SC & \cite{Moresco:2016mzx} \\ \hline
		cosmic\_clocks\_BC03 & cosmic clocks & $H(z)$ &  2.1 & SC & \cite{Moresco:2012jh} \\ \hline
		cosmic\_clocks\_MaStro & cosmic clocks & $H(z)$ &  2.1 & SC & \cite{Moresco:2012jh} \\ \hline		
		cosmic\_clocks\_BC03\_all & cosmic clocks &  $H(z)$ & 2.1  & SC & \cite{Moresco:2012jh} \\ 
		& & & & & \cite{Simon:2004tf} \\ 
		& & & & & \cite{Stern:2009ep} \\ \hline
		da\_rec & prior on angular diameter distance & $d_A(z_\mathrm{rec})$ & 1.1 & SC & \cite{Audren:2013nwa} \\ \hline
		gunn\_peterson & constraints on reionization history & $x_e(z)$ & 1.0 & SC & \cite{Fan:2006dp} \\ \hline
		hst & Hubble Space Telescope & $H_0$ prior & 3.0 & SC & \cite{Riess:2016jrr} \\ \hline
		igm\_temperature & constrains on baryon temperature & $T_b(z)$ & 1.0 & SC & \cite{Schaye:1999vr} \\ \hline
		ISW & NVSS,2MPZ,WI$\times$SC,SDSS/Planck & ISW & 3.0 & SC & \cite{Stolzner:2017ged} \\ \hline
		JLA & full JLA likelihood & Supernovae & 2.1 & D & \cite{Betoule:2014frx} \\ \hline
		JLA\_simple & simplified JLA likelihood & Supernovae & 2.1 & D & \cite{Betoule:2014frx} \\ \hline
		kids450\_qe\_likelihood\_public & KiDS-450 & Weak lensing & 3.0 & D & \cite{Kohlinger:2017sxk} \\ \hline
		lowlike & Planck 2013 + WMAP 9: low-$\ell$ & CMB & 1.2 & W & \cite{Ade:2013kta} \\ \hline
		Planck\_actspt & ACT 2013, SPT 2011 & CMB & 2.0 &W & \cite{2013JCAP...07..025D} \\ 
		& & & & & \cite{2011ApJ...743...28K} \\ \hline
		Planck\_highl & Planck 2015: TT high $\ell$ & CMB & 2.2 & W & \cite{Aghanim:2015xee} \\ \hline
		Planck\_highl\_lite & Planck 2015: TT high $\ell$ lite & CMB  & 2.2 & W & \cite{Aghanim:2015xee} \\ \hline
		Planck\_highl\_TTTEEE & Planck 2015: TTTEEE high $\ell$ & CMB & 2.2 & W & \cite{Aghanim:2015xee} \\ \hline
		Planck\_highl\_TTTEEE\_lite & Planck 2015: TTTEEE high $\ell$ lite & CMB & 3.0 & W & \cite{Aghanim:2015xee} \\ \hline
		Planck\_lensing & Planck 2015: lensing & CMB lensing & 2.2 & W & \cite{Ade:2015zua} \\ \hline
		Planck\_lowl & Planck 2015: TTTEEE low $\ell$ & CMB & 2.2 & W & \cite{Aghanim:2015xee} \\ \hline
		Planck\_SZ & Planck 2015: SZ cluster counts & Cluster Count & 2.2 & SC & \cite{Ade:2015fva} \\   
		& as $\Omega_m^\alpha \sigma_8$ prior & & &  & \\ \hline
		polarbear & Polarbear & CMB & 2.1 & SC & \cite{Ade:2014afa} \\ \hline
		quad & QUAD DR3 & CMB & 1.0 & SC & \cite{Brown:2009uy} \\ \hline
		sdss\_lrgDR4 & SDSS DR4: LRG & Galaxy Clust. & 3.0 & SC & \cite{Tegmark:2006az} \\ \hline
		sdss\_lrgDR7 & SDSS DR7: LRG & Galaxy Clust. & 3.0 & SC & \cite{Reid:2009xm} \\
		& & & & & \cite{Buen-Abad:2017gxg} \\ \hline
		simlow & from Planck 2016: TTTEEE low $\ell$ & $\tau_{\rm reio}$ prior & 3.0 & SC & \cite{Aghanim:2016yuo} \\
		& & & & & \cite{Adam:2016hgk} \\ \hline
		sn & Union2 & Supernovae & 1.0 & SC & \cite{Amanullah:2010vv} \\ \hline
		spt & SPT DR1 & CMB & 1.0 & SC & \cite{Schaffer:2011mz} \\ \hline
		spt\_2500 & SPT DR1, $\ell \leq 2500$ & CMB & 1.0 & SC & \cite{Schaffer:2011mz} \\ \hline
		timedelay & quasar time delays & Time Delay & 1.1 & SC & \cite{Suyu:2012aa} \\ \hline
		WiggleZ & WiggleZ power spectrum & Galaxy Clust. & 2.0 & SC & \cite{Parkinson:2012vd} \\ \hline
		WiggleZ\_bao & WiggleZ BAO & BAO & 2.1 & SC & \cite{Kazin:2014qga} \\ \hline
		wmap & WMAP 7yr (own wrapper) & CMB & 1.0 & D & \cite{Larson:2010gs} \\ \hline
		wmap\_9yr & WMAP 9yr (own wrapper) & CMB & 1.2 & D & \cite{Bennett:2012zja}
	\end{tabular}
	\caption{Current data likelihoods (letters c-z) \label{tab:likelihoods_data2}}
\end{table}

\begin{table}[t!]
	\begin{tabular}{ l | c | c | c | c | c }
		\multicolumn{6}{ c }{Forecast likelihoods} \\ %\hline
		name & description & type & LU & D & reference(s) \\ \hline \hline
		core\_m5 & CORE M5 ESA proposal & CMB & 3.0 & M & \cite{DiValentino:2016foa} \\
		& & & & & \cite{Finelli:2016cyd} \\ \hline
		euclid\_lensing & Euclid & Weak Lensing & 3.0$^*$ & M & \cite{Audren:2012vy} \\
		& & & & & \cite{Sprenger:2018tdb} \\ \hline
		euclid\_pk & Euclid & Galaxy Clust. & 3.0$^*$ & M & \cite{Audren:2012vy} \\
		& & & & & \cite{Sprenger:2018tdb} \\ \hline
		fake\_desi & DESI & BAO: $d_A/r_s$ & 3.0 & M & \cite{DiValentino:2016foa} \\ \hline
		fake\_desi\_euclid\_bao & best from DESI + Euclid & BAO & 3.0 & M & \cite{Font-Ribera:2013rwa} \\ \hline
		fake\_desi\_vol & DESI & BAO: $r_s/d_V$ & 3.0 & M & \cite{Archidiacono:2016lnv} \\ \hline
		fake\_planck\_bluebook & Planck 2015 est.: TTTEEE & CMB & 2.0 & M & \cite{Planck:2006aa} \\ \hline
		fake\_planck\_realistic & Planck 2018 est.: TTTEEE$\phi\phi$ & CMB & 3.0 & M & \cite{DiValentino:2016foa} \\ \hline
		litebird & LiteBIRD est. & CMB & 3.0 & M & \cite{DiValentino:2016foa} \\
		& & & & & \cite{Suzuki:2018cuy}\\ \hline
		ska1\_IM\_band1 & SKA1 band 1 & 21cm Int. Map. & 3.0$^*$ & M & \cite{Sprenger:2018tdb} \\ \hline
		ska1\_IM\_band2 & SKA1 band 2 & 21cm Int. Map. & 3.0$^*$ & M & \cite{Sprenger:2018tdb} \\ \hline
		ska1\_lensing & SKA1 & Weak Lensing & 3.0$^*$ & M & \cite{Sprenger:2018tdb} \\ \hline
		ska1\_pk & SKA1 & Galaxy Clust. & 3.0$^*$ & M & \cite{Sprenger:2018tdb} \\ \hline
		ska2\_lensing & SKA2 & Weak Lensing & 3.0$^*$ & M & \cite{Sprenger:2018tdb} \\ \hline
		ska2\_pk & SKA2 & Galaxy Clust. & 3.0$^*$ & M & \cite{Sprenger:2018tdb}
	\end{tabular}
	\caption{Forecast likelihoods.
	$^*$ Euclid likelihoods will be updated and SKA likelihoods published when the relevant publication has been accepted for publication.\label{tab:likelihoods_forecast}}
\end{table}

\bibliographystyle{elsarticle-num}
\biboptions{numbers,sort&compress}
\bibliography{references}

\end{document}